\documentclass[twocolumn,showpacs,pra]{revtex4-1}
\usepackage{amsmath}
\usepackage{amssymb,amsfonts}
\usepackage{amsxtra}
\usepackage[mathscr]{eucal}
\usepackage{epsf}
\usepackage[usenames,dvipsnames]{color}
\usepackage{graphicx}
\usepackage{dcolumn}
\usepackage{bm}
\usepackage{esint}
\begin{document}
\title{Nanoscopy of pairs of atoms by fluorescence in a magnetic field}

\author{E. S. Redchenko$^{1,2}$, A. A. Makarov$^{1,3}$, and V. I. Yudson$^{4,1}$}
 \affiliation{
$^{1}$Institute of Spectroscopy, Russian Academy of Sciences,
5 Fizicheskaya St., Troitsk, Moscow 108840, Russia \\
 $^{2}$Institute of Science and Technology Austria, 3100 Klosterneuburg, Austria\\
 $^{3}$Moscow Institute of Physics and Technology, Institutskiy
 pereulok 9, 
Dolgoprudny, Moscow Region 141700, Russia \\
$^4$Laboratory for Condensed Matter Physics, National Research 
University 
Higher School of Economics, 20 Myasnitskaya St., Moscow 101000, Russia}
\date{\today}
\begin{abstract}
Spontaneous emission spectra of two initially excited closely spaced identical atoms are 
very
sensitive to the strength and the direction of the applied magnetic field. We consider
the relevant schemes 
that ensure the determination of the mutual spatial orientation of the atoms
and the distance between them by entirely optical means. A corresponding theoretical description 
is given accounting for the dipole-dipole interaction between the two atoms in the presence of 
a magnetic field and for polarizations of the quantum field interacting with magnetic sublevels 
of the two-atom system.
\end{abstract}

\maketitle
\section{INTRODUCTION}
Ultrahigh spatial resolution at the distances short compared to optical wavelength is
a challenging  spectroscopic problem. Various approaches to solution of this problem
have been suggested, and some of them  have been realized in the experiment.
The following ideas can be mentioned: (i) the photoelectron (photoion) microscopy
\cite{LETOKHOV1,LETOKHOV2,LETOKHOV3} being a development of the M\"{u}ller emission 
microscope \cite{MULLER}; (ii) the transport of the molecule excitation between a 
nanosize needle tip and bulk via the
fluorescence resonance energy transfer (FRET) \cite{KOPELMAN};
(iii)
FRET scanning near-field optical microscopy \cite{SL,SEKATSKII}; (iv) combinations of the optical excitation with nano-resolution of the atomic force microscope or scanning tunnel microscope \cite{LAPSHIN};
(v) the transport of a laser excited atom through a nanohole in a metal screen \cite{BALYKIN}.

A system of two closely spaced identical atoms (quantum dots, vacancy centers, organic molecules in solid solutions, etc.) is the simplest model where a determination of, at least, two parameters might be desirable. These parameters are the distance between the atoms
and the direction of their vector separation in the space. In this paper, we consider a 
method for their estimation by only optical means, {\textit{without any nano-tools}}
like needles, tips, or holes. (For a possible solution of this task in 2D see, e.g., Ref.~\cite{SUN}).

A system of two identical atoms with the ground states $|g_{1,2}\rangle$ and the excited states $|e_{1,2}\rangle$ is an example exhibiting the effects of Dicke super- and subradiance \cite{DICKE}. The two excited
states of this system are $|{\mathcal{Q}}_1\rangle=|e_1g_2\rangle$ and
$|{\mathcal{Q}}_2\rangle=|g_1e_2\rangle$. The
dipole-dipole interaction between the states $|{\mathcal{Q}}_1\rangle$ and
$|{\mathcal{Q}}_2\rangle$ leads to their
superpositions producing two {\em{entangled}} eigenstates, symmetric and antisymmetric
(see, e.g., Refs. \cite{FICEK,BGZ}). While the symmetric state $|{\mathcal{Q}}_s\rangle=
{\frac{1}{{\sqrt{2}}}}\left(|{\mathcal{Q}}_1\rangle+|{\mathcal{Q}}_2\rangle\right)$ is superradiant, decaying twice faster than the one-atom state $|e\rangle$, the
antisymmetric state $|{\mathcal{Q}}_a\rangle=
{\frac{1}{{\sqrt{2}}}}\left(|{\mathcal{Q}}_1\rangle-|{\mathcal{Q}}_2\rangle\right)$ is subradiant, decaying slowly by the parameter $(r/\lambda_{eg})^2$ if the distance
$r$ between the atoms is smaller then the wavelength of the $|e\rangle\rightarrow|g\rangle$
transition.
The theory of this system (with an emphasis on superradiance) was studied in a number of
works (see, e.g., Refs. \cite{LEMBERG,MILLONI,WANG}, and the review \cite{FICEK}). An interesting effect of the exchange between symmetric and antisymmetric states due to the spatial variation of the applied laser pulse at the positions of the atoms was discovered in Ref. \cite{DAS}. As for the subradiant states, they attracted a special attention because of their property of a relatively slow spontaneous decay and, hence, a potentiality of keeping quantum information for a long time.

A prominent effect of subradiance can be achieved in the one-dimensional (1D) case \cite{ML,RY} (using, e.g., a single-mode waveguide or a photon crystal) where two spatially separated atoms are placed at a distance of the whole number of the half-wavelengths of the optical transition. Several methods were suggested to produce subradiant states. E.g., in the mentioned 1D case, the subradiant state is produced with a probability of
$\approx1/2$, when, at the initial moment of time, one of the atoms is excited. 
More complicated configuration in a 1D waveguide have been also considered (see, e.g. 
\cite{Bar-⁠2017}). However, this schemes can be realized only for spatially separated atoms. For closely spaced atoms in 3D, two schemes of control of subradiance can be
mentioned \cite{ZADKOV,JETPLETT}, the both showing reasonable fidelity.
An interesting scheme for producing the single-photon subradiant state in an ensemble of many atoms was recently considered by Scully \cite{SCULLY1}. Protection of this state was 
discussed in Ref. \cite{SCULLY2}.
As for the experiment on subradiance, certain evidences for modification of the decay rate 
were received, e.g., in Ref.~\cite{PAVOLINI} for large atomic ensemble, and in 
Ref.~\cite{BREWER} for a system of two trapped ions. Subradiance
was also observed very recently for a cloud of cold atoms \cite{GUERIN}. The 
contribution of this effect was small, but detectable as a
narrow spectrum of fluorescence gated at times much longer than the time of single-atom 
decay.
In connection with this experimental work, a recent theoretical paper 
\cite{HKOR} should be mentioned where the property of subradiance in atomic ensembles has 
been argued to be more general than this is usually assumed.

\begin{figure}[t]
\includegraphics[width=0.5\textwidth]{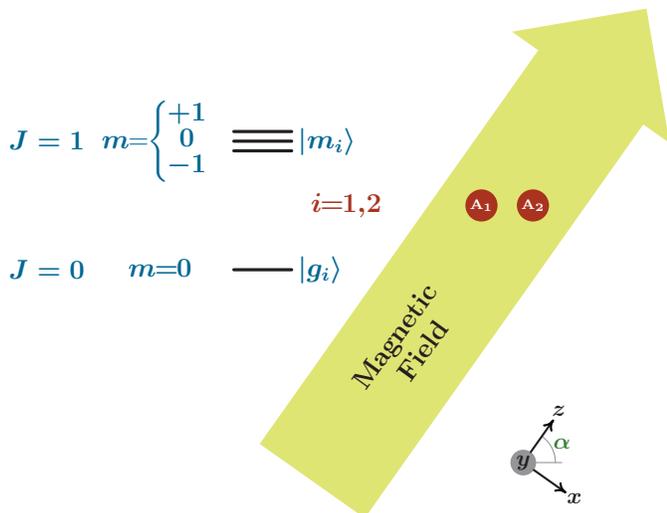}
\caption{Our model. Two identical atoms A$_1$ and A$_2$ are located in the $xz$ plane. The angular momenta 
of their ground state $|g\rangle$ and excited state $|e\rangle$ are, respectively, $J_g=0$
and $J_e=1$. Magnetic field is 
directed along the $z$ axis. The line connecting the atoms is
directed at a certain angle $\alpha$ to the z axis. This leads to splitting of the excited 
state into the triplet of states with the $J$-projections $m=0,\pm1$ on the $z$ axis.
(In our consideration, the main results refer to the magic angle 
$\alpha=\alpha_{\textrm{mag}}=\arccos(1/{\sqrt{3}})\approx54.7^o$.)}
\label{model_figure_m}
\end{figure}
\begin{figure}[t]
\centering
\includegraphics[width=0.5\textwidth]{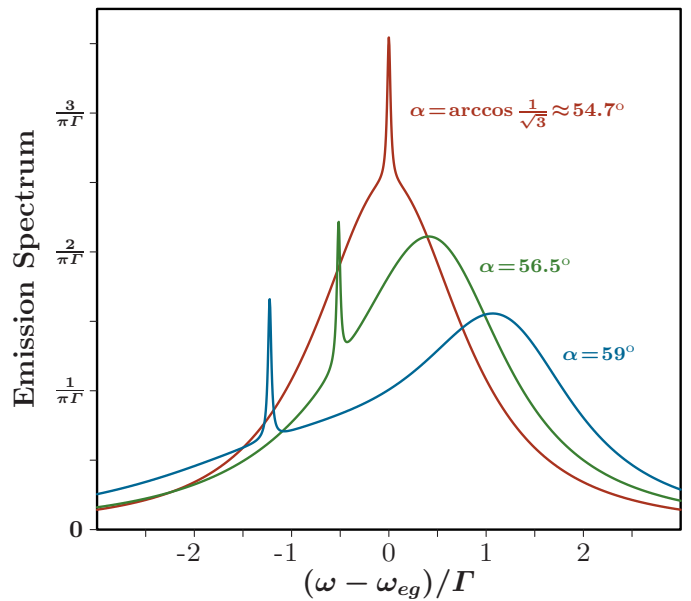}
\caption{An example of emission spectra of a two-atom system in
a strong magnetic field ${\textbf{H}}$ at different values of the angle $\alpha$ (see
Fig.~\ref{model_figure_m}). Initially 
(at $t=0$) the both atoms are excited to the state $|e\rangle$ with $m=0$.
The rate of spontaneous transition $|e\rangle\rightarrow|g\rangle$ is $\varGamma$. 
The transition frequency is $\omega_{eg}$ and the
corresponding wave vector is $k_0$. For this example, the atoms are located so that 
$k_0r=0.5$. For calculation of these spectra, see Sec.~\ref{angular-resolution}.} 
\label{spectrum-strong-H}
\end{figure}

In this paper we consider the fluorescence spectra of two closely spaced identical 
atoms in a magnetic field. We use the remarkable fact that, being directed at the magic 
angle $\alpha\approx\arccos(1/\sqrt{3})\approx54.7^{\textrm{o}}$ relative to the 
direction from one atom to another (see Fig. \ref{model_figure_m}), a sufficiently strong 
magnetic field suppresses the dipole-dipole excitation transfer from one atom to the other
one \cite{CONTROL}. This makes the fluorescence spectrum
following the excitation of the both atoms be {\em{very sensitive to small 
deflections from the magic-angle configuration}} (see Fig. \ref{spectrum-strong-H}). 
This observation helps 
one to determine the direction of the vector separation by a series of manipulations with a 
magnetic field. To demonstrate the relevant 
procedure, first, in Sec.~II, we describe the model, write down the Hamiltonian of our 
system, and derive properties of its eigenstates. Then, in Sec. III, we consider general
theory of 
interaction of the two-atom system with the electromagnetic field. Next, in Sec.~IV, we 
consider its consequences concerning fluorescence spectra in a magnetic field 
of (i) singly-excited entangled states (Sec.~\ref{Decay-singly}), and (ii) doubly-excite 
states Sec.~\ref{Decay-doubly}) where, in particular, it is shown 
(Sec.~\ref{Reach-antisymm}) how the distance between 
the atoms can be estimated from the probability of the gated fluorescence 
from the subradiant state only.
Finally, in Sec. V, 
we show how the 
{\em{full nanoscopy}} of a pair of atoms can be 
completed, i.e. in addition to the distance between the atoms 
(Sec.~\ref{Reach-antisymm}), the direction of the vector separation from one atom to 
another can be determined. The latter is achieved using 
the above announced sensitivity (see Fig. \ref{spectrum-strong-H} supported by calculations 
of Sec.~\ref{angular-resolution}) of the fluorescence spectra to the direction of the 
magnetic field.

The calculation details are presented in the Appendices.   
\section{Description of the two-atom system}
The system of two closely located identical atoms in an external magnetic field $\bf{H}$ directed along the
$z$-axis is shown in Fig. {\ref{model_figure_m}}.
 The vector $\bf{r}$ connecting the positions of
two atoms meets at an angle $\alpha$ with the field $\bf{H}$; below we choose the 
$x$ and $y$ coordinate axes such that
\begin{equation} 
{\bf{r}} =(r_x, r_y, r_z) = r(\sin{\alpha}, 0, \cos{\alpha})\;.
\end{equation}
It is assumed that the ground state $|g\rangle$ of the atom is non-degenerate and corresponds to the angular momentum $J=0$, while the excited state $|e\rangle$ of our
interest  is a triplet of states $|e^{(m)}\rangle$ (just 
$|m\rangle$ below for shortness) corresponding to the momentum $J=1$ with projections 
$m=0, \pm 1$ on the quantization axis $z$. 
In a zero magnetic field this triplet is degenerate
and its energy (counted from the ground atomic state) is $\omega_{eg}$. Nonzero components of the dipole moment operator $\widehat{{\bf{d}}}$ between the states $|g\rangle$ and 
$|m\rangle$ are (see Appendix A) given by
\begin{equation}\label{dipole-moments}
\langle m=0|\hat{d}_z|g\rangle \equiv d_0\;, \,\,\,\,
\langle m=\pm 1|\hat{d}_{\pm}|g\rangle = \sqrt{2}d_0 \; ,
\end{equation}
where $\hat{d}_{\pm}\equiv \hat{d}_x \pm i\hat{d}_y$ are the circular components.

The reduced Hilbert space of the two-atom system can be described by the following basis:
$\left\{ |g,g\rangle; |m,g\rangle; |g,m\rangle, |m,m'\rangle \right\}$, where the
two-atom state $|a, b\rangle = |a\rangle_1\otimes |b\rangle_2$ is a direct product
of the states of the first and the second atoms.
The first state in curly brackets corresponds to the ground state, the 
second and third ones---to singly excited states, and the forth one to a doubly excited
state of the two-atom system. 
In this paper we consider only symmetric doubly-excited states of the form $|m,m\rangle$ 
that can be created by an optical $\pi$-pulse of a certain polarization, linear or 
circular. 
To account for the symmetry of the two-atom system with one excitation, it is convenient to introduce new basis of symmetric and antisymmetric states:
\begin{eqnarray}
&&|s_m\rangle = \frac{1}{\sqrt{2}}(|m,g\rangle+|g,m\rangle)\; ,  \label{S-state}\\
&&|a_m\rangle = \frac{1}{\sqrt{2}}(|m,g\rangle-|g,m\rangle)\; \label{A-state}.
\end{eqnarray}
The Hamiltonian $\widehat{H}$ of the two atom system can be represented as a sum
$\widehat{H}^{(1)} + \widehat{H}^{(2)}$ of two Hamiltonians acting in the subspaces with one and two
excitations, respectively. An important constituent of the Hamiltonian $\widehat{H}$ of closely
located atoms is the operator $\widehat{U}$ of the dipole-dipole interatomic interaction:
\begin{eqnarray}\label{U}
\widehat{U}=\frac{\left(\widehat{\bf{d}}_1\!\cdot\!\widehat{\bf{d}}_2\right)}{r^3}-
3\frac{\left(\widehat{\bf{d}}_1\!\cdot\!\bf{r}\right)\left(\widehat{\bf{d}}_2\!\cdot\!
\bf{r}\right)}{r^5} \; ,
\end{eqnarray}
where $\widehat{\bf{d}}_{1(2)}$ is the dipole moment operator of the first (second) atom. This interaction is not important for doubly excited states: the induced coupling between doubly excited and ground states results only in a parametrically weak 
($\sim d^2_0/\hbar r^3\omega_{eg} \ll 1$) renormalization of state parameters 
that will be neglected below. So, the
Hamiltonian $\widehat{H}^{(2)}$ in the subspace of our interest can be represented as
\begin{eqnarray}\label{H2}
\widehat{H}^{(2)}= \sum_m (2\hbar\omega_{eg} + 2m\varOmega_H)|m,m\rangle\langle m.m|
 \; ,
\end{eqnarray}
where $\varOmega_H = \mu_{\textrm{B}}g H $ describes the Zeeman splitting of levels in the magnetic field ($\mu_{\textrm{B}}$ is the Bohr 
magneton, and g is the Land\'{e} factor).

The interaction $\widehat{U}$ is mixing the states with different 
projections $m$ in the subspace with one excitation. As $\widehat{U}$ does not mix states 
with different symmetries, the matrix of the Hamiltonian 
$\widehat{H}^{(1)}$ in the basis (\ref{S-state})--(\ref{A-state}) is split 
into two parts, $\widehat{H}^{(1)}_s$ and 
$\widehat{H}^{(1)}_a$:
\begin{widetext}
\begin{eqnarray}\label{Matrix}
&& \widehat{H}^{(1)}_s = \hbar\omega_{eg} \widehat{\mathbb{I}} +
U_r\begin{pmatrix} -\frac{\varOmega_H}{U_r}\!+\!(1\!-\!\frac{3}{2}\sin^2\alpha) & 
-\frac{3\sqrt{2}}{4}\sin2\alpha & -\frac{3}{2}\sin^2\alpha \\ -\frac{3\sqrt{2}}{4}\sin2\alpha & (1\!-\!3\cos^2\alpha) & -\frac{3\sqrt{2}}{4}\sin2\alpha \\ -\frac{3}{2}\sin^2\alpha & -\frac{3\sqrt{2}}{4}\sin2\alpha & \frac{\varOmega_H}{U_r}\!+\!(1\!-\!\frac{3}{2}\sin^2\alpha) \end{pmatrix}\; ,\\
&& \widehat{H}^{(1)}_a =\hbar\omega_{eg} \widehat{\mathbb{I}} + U_r \begin{pmatrix} -\frac{\varOmega_H}{U_r}\!-\!(1\!-\!\frac{3}{2}\sin^2\alpha) & \frac{3\sqrt{2}}{4}\sin2\alpha & \frac{3}{2}\sin^2\alpha \\ \frac{3\sqrt{2}}{4}\sin2\alpha & -(1\!-\!3\cos^2\alpha) & \frac{3\sqrt{2}}{4}\sin2\alpha \\ \frac{3}{2}\sin^2\alpha & \frac{3\sqrt{2}}{4}\sin2\alpha & \frac{\varOmega_H}{U_r}\!-\!(1\!-\!\frac{3}{2}\sin^2\alpha) \end{pmatrix}\;,
\label{Matrix-a}
\end{eqnarray}
\end{widetext}
where the rows and columns are enumerated in the order $m=-1, 0, 1$, 
$\widehat{\mathbb{I}}$ is the unit matrix in the space spanned by 
(\ref{S-state})--(\ref{A-state}), $U_r=d_0^2/ r^3$, and an explicit form of angular 
parts of the wavefunctions corresponding to $|g\rangle$ and $|m\rangle$ is  
given in Appendix A. 

\begin{figure}[t]
\centering
\includegraphics[width=0.5\textwidth]{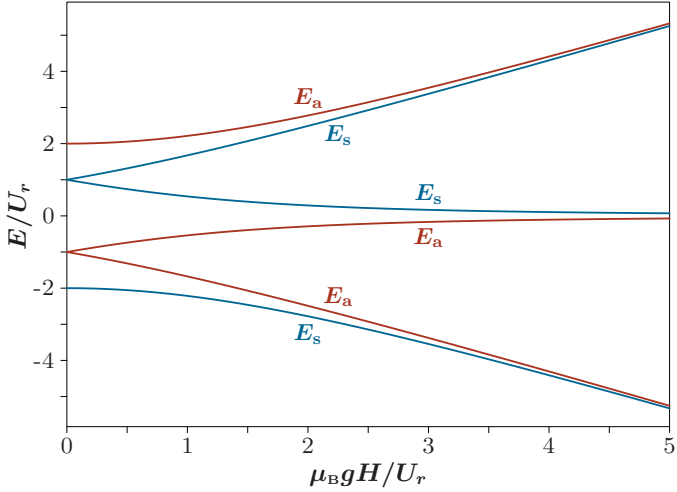}
\caption{Energies of eigenstates of a system of two atoms as functions of the magnetic 
field ${\bf{H}}$ applied at the magic angle as shown in Fig.~\ref{model_figure_m}.}
\label{allenergy_figure}
\end{figure}

Eqs. (\ref{Matrix})--(\ref{Matrix-a}) are simplified  for 
the above introduced magic angle 
(see Fig.~\ref{model_figure_m}), when the 
diagonal part of the dipole-dipole interaction vanishes, and have next form 
\begin{eqnarray}\label{Matrix_magic}
&& \widehat{H}^{(1)}_s =\hbar \omega_{eg} \widehat{\mathbb{I}} + U_r\begin{pmatrix} -\frac{\varOmega_H}{U_r} & -1 & -1 \\ -1 & 0 & -1 \\ -1 & -1 & \frac{\varOmega_H}{U_r} \end{pmatrix}\, , \\
&& \widehat{H}^{(1)}_a = \hbar\omega_{eg} \widehat{\mathbb{I}} + U_r\begin{pmatrix} -\frac{\varOmega_H}{U_r} & 1 & 1 \\ 1 & 0 & 1 \\ 1 & 1 & \frac{\varOmega_H}{U_r} \end{pmatrix}\, .
\label{Matrix-a_magic}
\end{eqnarray}
The eigenenergies of the two Hamiltonians are determined by the characteristic equations:
\begin{eqnarray}\label{SA}
\left(E^{s(a)}\right)^3-\left(\varOmega_H^2+3U_r^2\right)E^{s(a)}\pm 2U_r^3=0\;,
\end{eqnarray}
where $E^{s(a)}$ are counted from the atomic transition frequency $\omega_{eg}$.
In the absence of the magnetic field, the solutions $E^{s(a)}_j$ of the characteristic 
equations (it is convenient to numerate them by $j=-1,0,1$
in the ascending order) are:
\begin{equation}\label{No_field}
E^s_{-1}\!=\!-2U_r,\;  E^s_{0(1)}\!=\!U_r ; \; E^a_{-1(0)}\!=\!-U_r,\;E^a_{+1}\!=\!2U_r \; .
\end{equation}
In the opposite case of a strong magnetic field, the role of the dipole-dipole interaction is small, the eigenstates only slightly differ from $|s_m\rangle$ and 
$|a_m\rangle$, and eigenenergies are given by
\begin{eqnarray}
 E^\nu_{\pm 1}&& \approx m\left(\varOmega_H + \frac{3U_r^2}{2\varOmega_H}\right)
-\frac{\nu U_r^3}{\varOmega_H^2}
 \label{High-H-pm1}\;,\\
 E^\nu_{0} &&\approx \frac{2\nu U_r^3}{\varOmega^2_H}
\label{High-H-m0}\; ,
\end{eqnarray}
where $\nu = \pm 1$ for the symmetric and antisymmetric states, respectively.
For an arbitrary magnitude of the magnetic field (and $\alpha = \alpha_{\textrm{mag}}$), the eigenenergies are presented in Fig. {\ref{allenergy_figure}}.

For further purposes we also present
the expression for 
the energy  $E^\nu_{0}$ for the case of an arbitrary angle $\alpha$:
\begin{equation}\label{High_field non-mag}
E^\nu_{0} \approx \nu U_r \left[1- 3\cos^2{\alpha}+{\frac{27U_r^2}{\varOmega_H^2}}\sin^2{\alpha}\cos^4{\alpha}\right] \; .
\end{equation}
In a close vicinity of the magic angle this expression takes the form:
\begin{eqnarray}\label{High_field near-mag}
E^\nu_{0} \approx2\nu U_r\left[{\sqrt{2}}(\alpha - \alpha_{\textrm{mag}})+ 
\frac{U_r^2}{\varOmega_H^2} \right]\; .
\end{eqnarray}
As seen from Eqs. (\ref{High-H-m0})--(\ref{High_field near-mag}), the detuning
of $E^\nu_{0}$ from zero decreases dramatically when 
$\alpha\rightarrow\alpha_{\textrm{mag}}$.
The same is for the splitting between $E^s_{\pm1}$ and $E^a_{\pm1}$. 

Eigenvectors $|\psi_j^s\rangle$ and $|\psi_j^a\rangle$ (numerated by $j=-1,0,1$)
of the matrices $\widehat{H}^{(1)}_s$ and $\widehat{H}^{(1)}_a$
form bases connected with $\left\{|s_m\rangle \right\}$ and
$\left\{|a_m\rangle \right\}$ by a unitary matrix $\widehat{C}^{\nu}$
\begin{eqnarray}\label{C-matrix}
|\nu_m\rangle=\sum_{j}C_{mj}^{\nu}|\psi_j^{\nu}\rangle\; ,
\end{eqnarray}
where $\nu = s, a$. These eigenvectors can be expressed in terms
of the elements of the matrices (\ref{Matrix_magic})--(\ref{Matrix-a_magic}),
and the corresponding eigenenergies 
$E^\nu_{j}$ ($\nu = s,a$; $j=-1,0,1$) defined by Eq.~(\ref{SA}).
In the basis $|\nu_m\rangle$ ($m =-1, 0, 1$)
defined by Eqs. (\ref{S-state})--(\ref{A-state})
they are
\begin{eqnarray}\label{Eig_basis}
|\psi_j^{\nu}\rangle = q^{\nu}_{j}
\left(
  \begin{array}{c}
    \frac{E^\nu_{j}\mp U_r}{
    \varOmega_H+E^\nu_{j} \mp U_r}    \\
     1 \\
     - \frac{E^\nu_{j} \mp U_r}{\varOmega_H- E^\nu_{j} \pm U_r} \\
  \end{array}
\right)
\equiv \sum_{m}Q^{\nu}_{jm}|\nu_m\rangle\; ,
\end{eqnarray}

where the upper (lower) sign refers to the symmetric (antisymmetric) state $s$ ($a$), and
\begin{equation}\label{norma}
q^{\nu}_{j} \!= \!\left[\left( \frac{E^\nu_{j}\mp U_r}
    {\varOmega_H\!+\!E^\nu_{j} \mp U_r}\right)^2 \!   + \!1
 \! \right. \left.   +\!\left(  \frac{E^\nu_{j} \mp U_r}{\varOmega_H\!-\! E^\nu_{j} 
 \pm \!U_r\!}\right)^2 \right]^{-1/2}
  \, .
\end{equation}

The matrix $\widehat{C}^{\nu}$ is connected with $\widehat{Q}^{\nu}$ as $C^{\nu}_{mj}=Q^{\nu}_{jm}$.

In the next section we develop a convenient description of the interaction
of the two-atom system with the electromagnetic field.

\section{Interaction with the electromagnetic field}
The Hamiltonian of the two-atom system interacting with the electromagnetic
field has the form:
\begin{eqnarray}\label{H-cal}
&&\widehat{\mathcal{H}}=\widehat{H} + \widehat{H}_{\textrm{ph}}+
\widehat{H}_{\textrm{int}} \; ,\\
&&\widehat{H}_{\textrm{ph}}=\sum_{{\bf k}\lambda}\hbar\omega_ka_{{\bf k}
\lambda}^{\dag}a_{{\bf k}\lambda}
\; ,
\end{eqnarray}
where the two-atom Hamiltonian $\widehat{H}$ has been defined in the previous 
section, and
$\widehat{H}_{\textrm{ph}}$ corresponds to free 
photons, 
$\omega_k$ and $\bf{k}$ being the photon frequency and wave vector, so 
$a_{{\bf k}\lambda}^{\dag}$ ($a_{{\bf k}\lambda}$) is a photon creation (annihilation) operator with  $\lambda=1,2$ meaning the two linear polarizations.
The interaction Hamiltonian $\widehat{H}_{\textrm{int}}$ can be split into two parts 
$\widehat{H}^{(1)}_{\textrm{int}}+\widehat{H}^{(2)}_{\textrm{int}}$ 
that correspond to optical transitions
between the ground and a singly excited state of the two-atom system and
between the singly and a doubly excited states, respectively.

In the rotating wave approximation (RWA) the first part, $\widehat{H}^{(1)}_{\textrm{int}}$,
can be represented as
\begin{eqnarray}\label{H-int-1}
&& \widehat{H}^{(1)}_{\textrm{int}}=-\sum_{{\bf{k}},\lambda, m} \left\{ g_{\bf{k}\lambda}^{m,g}
\left [ |m,g\rangle\langle g,g| e^{-i\frac{\bf{k}\bf{r}}{2}} \right. \right. \nonumber \\
&& \left. \left. +
|g,m\rangle\langle g,g|e^{i\frac{\bf{k}\bf{r}}{2}}\right] a_{\bf{k}\lambda} + {\mathrm{H.c.}}\right\} \; .
\end{eqnarray}
Here H.c. means Hermitian conjugate,
the first and the second terms in the square bracket correspond, respectively, to the 
optical excitation of only the first atom 
(located at $-{\bf{r}}/2$) and of only the second
atom (located at ${\bf{r}}/2$), the matrix element of the dipole transition is given by
\begin{equation}\label{g}
g_{{\bf {k}}\lambda}^{m,g}= \sqrt{\frac{2\pi\hbar\omega_{{\bf k}}}{V}}
\langle m|{\hat{{\bf{d}}}}|g\rangle
{\bf{e}}_{{\bf {k}}\lambda} \, ,
\end{equation}
where $V$ is the quantization volume, $\bf{e}_{{\bf k}\lambda}$ is the polarization
vector. Below we will replace the photon frequency $\omega_{{\bf k}}$ in Eq. (\ref{g}) by 
the resonance transition frequency $\omega_{eg}$.
A similar expression can be written for the part $\widehat{H}^{(2)}_{\textrm{int}}$ with the only difference that the transitions take place between the states $|m,g\rangle$ (or $|g,m\rangle$) and $|m,m'\rangle$ ($|m',m\rangle$). Restricting our consideration to only symmetric doubly excited states $|m,m\rangle$,
we have
\begin{eqnarray}\label{H-int-2}
&& \widehat{H}^{(2)}_{\textrm{int}}=-\sum_{\bf{k},\lambda, m} \left\{ g_{\bf{k}\lambda}^{m,g}
|m,m\rangle
\left[\langle m, g| e^{i\frac{\bf{k}\bf{r}}{2}} \right. \right. \nonumber \\
&& \left. \left. +
\langle g,m|e^{-i\frac{\bf{k}\bf{r}}{2}}\right] a_{\bf{k}\lambda} + {\mathrm{H.c.}}\right\} \, .
\end{eqnarray}
The choice of the polarization vectors and explicit expressions for the matrix element 
given by Eq. (\ref{g}) are described in the Appendix A. Here we only briefly discuss the main steps of the further analysis.

(i) We express singly excited states of the 
two-atom system in terms of the symmetric and antisymmetric combinations (\ref{S-state}) 
and (\ref{A-state}), and,  finally, in terms of exact states
$\left\{\right |\psi^{\nu}_j \rangle; j=-1,0,1\}$ [see Eq. (\ref{C-matrix})] of a definite
symmetry ($\nu = s, a$).

(ii) To simplify calculations we use the symmetry of the system and introduce new
(``symmetric'' and ``antisymmetric") photon operators:
\begin{eqnarray}
&& b_{1s}(\bf{k}) = \frac{1}{\sqrt{2}}[a_{\bf{k}1} - a_{-\bf{k}1}], \label{b1s}\\
&& b_{1a}(\bf{k}) = \frac{1}{\sqrt{2}}[a_{\bf{k}1} + a_{-\bf{k}1}], \label{b1a}\\
&& b_{2s}(\bf{k}) = \frac{1}{\sqrt{2}}[a_{\bf{k}2} + a_{-\bf{k}2}], \label{b2s} \\
&& b_{2a}(\bf{k}) = \frac{1}{\sqrt{2}}[a_{\bf{k}2} - a_{-\bf{k}2}], \, .\label{b2a}
\end{eqnarray}

To avoid double counting we restrict the photon wave vector $\bf{k}$ to the
upper hemisphere $k_z\geqslant0$. Then the above set of new operators
is complete, and obeys the usual boson commutation relations:
\begin{eqnarray}\label{b-c-commutators}
&&[b_{1s}({\bf{k}}),b^{\dagger}_{1s}({\bf{k}}')]=\delta_{{\bf{k}},{\bf{k}}'}
 =[b_{2s}({\bf{k}}),b^{\dagger}_{2s}({\bf{k}}')]\;,\\
&&[b_{1s}({\bf{k}}),b^{\dagger}_{1a}({\bf{k}}')]=0 \;.
\end{eqnarray}
Now, the free photon Hamiltonian is 
\begin{equation}\label{H-photon}
\widehat{H}_{\textrm{ph}}=\sum\nolimits^\prime_{{\bf k}; \lambda=1,2; \nu =s,a}
\hbar\omega_k
b_{\lambda \nu}^\dagger({\bf{k}})
b_{\lambda \nu}({\bf{k}})
\;,
\end{equation}
where the prime sign denotes the summation over the upper hemisphere of $\bf{k}$.
The interaction part becomes
\begin{multline}
\widehat{H}^{(1)}_{\textrm{int}}=-2\sideset{}{'}\sum_{{\bf{k}}'\!,\lambda,j}
\left\{g_{{\bf{k}}'\!\lambda}^{m,g}\left[\cos\left({\frac{{\bf{kr}}}{2}}\right)
C_{mj}^{s}|\psi_j^s\rangle b_{\lambda s}({\bf{k}}')\right.\right. \\
\left.\left. -i\sin\left({\frac{{\bf{kr}}}{2}}\right)C_{mj}^{a}
|\psi_j^a\rangle b_{\lambda a}({\bf{k}}')\right]\langle g,g|  + {\mathrm{H.c.}} \right\} \, ,
\label{H-int-1-b} 
\end{multline}
\begin{multline}
\widehat{H}^{(2)}_{\textrm{int}}=-2\sideset{}{'}\sum_{{\bf{k}}'\!,\lambda,j}
\left\{g_{{\bf{k}}'\!\lambda}^{m,g}|m,m\rangle\left[\cos\left({\frac{{\bf{kr}}}{2}}\right)
C_{mj}^{s}|\psi_j^s\rangle b_{\lambda s}({\bf{k}}')\right.\right. \\
\left.\left. +i\sin\left({\frac{{\bf{kr}}}{2}}\right)C_{mj}^{a}
|\psi_j^a\rangle b_{\lambda a}({\bf{k}}')\right]  + {\mathrm{H.c.}} \right\} \, ,
\label{H-int-2-b} 
\end{multline}
where the matrices $C_{mj}^{\nu}$ ($\nu = s, a$) are defined by Eq.~(\ref{C-matrix}).

\begin{figure}[t]
\centering
\includegraphics[width=0.5\textwidth]{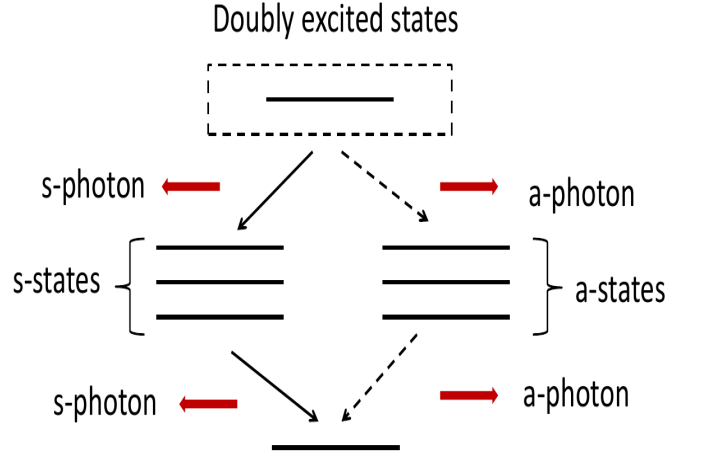}
\caption{Scheme of the spontaneous decay of
the doubly excited states of a system of two atoms. There are two ways of the decay whose 
branching ratio depends on the distance between the atoms.}
\label{Fig-decay}
\end{figure}

Eqs. (\ref{H-photon})--(\ref{H-int-2-b}) explicitly demonstrate the important properties of the system:
(i) the symmetric and antisymmetric singly excited states decay to the ground states
independently of each other; the decay is caused by the interaction with only
symmetric or antisymmetric photon continua, respectively;
(ii) the decay of a doubly excited state (of 
the form $|m,m\rangle$) to a
singly excited one can go both via the symmetric and antisymmetric channels.
The decay scheme is shown in Fig.~\ref{Fig-decay}.

In the case of interest, when the interatomic separation is small ($kr \ll 1$), the matrix elements of the symmetric and antisymmetric channels enter
Eqs. (\ref{H-int-1}) and (\ref{H-int-2}) with rather different weights:
the symmetric one $\propto \cos\left(\frac{\bf{kr}}{2}\right)\approx 1$, while the antisymmetric one $\propto \sin\left(\frac{\bf{kr}}{2}\right)\approx
{\bf{kr}}/2 \ll 1$. This means that the ratio of the decay rates via corresponding channels scales as $\varGamma_a/\varGamma_s \propto (k_0r)^2 \ll 1$, where $k_0$ is
the wave vector that corresponds to the transition frequency $\omega_{eg}$.

Some implementations  of the developed formalism are given in the subsequent sections.

\section{Fluorescence spectra of the two-atom system}
\subsection{Decay rate of a singly excited system}\label{Decay-singly}
Consider the two-atom system originally (at the time $t=0$)
prepared in one of the eigenstates $\psi^{\nu}_j$ of a definite
symmetry $\nu = s,a$. The time evolution of the atom-photon system is described by
the wave function
\begin{multline}\label{Psi1-t}
|\varPsi(t)\rangle = A_j^{\nu}(t)|\psi^{\nu}_j\rangle\otimes|{\textsl{vac}}\rangle\\
+\sideset{}{'}\sum_{{\bf{k}},\lambda}
B^{\nu}_{{\bf{k}}\lambda}(t)|g,g\rangle\otimes b_{\lambda \nu}^{\dagger}(\bf{k})
|{\textsl{vac}}\rangle
\,
\end{multline}
determined by the Schr\"{o}dinger equation with the interaction Hamiltonian $\widehat{H}^{(1)}_{\textrm{int}}$ (\ref{H-int-1-b}) and the initial condition $A_j^{\nu}(0)=1$,
$B^{\nu}_{{\bf{k}}\lambda}(0)=0$; 
the state $|{\textsl{vac}}\rangle$ is the vacuum state
of the electromagnetic field (no photons). Assuming that the interval between
energy levels of the same symmetry is much greater than the level's widths
we have neglected in Eq.~(\ref{Psi1-t}) an admixture of other excited states
in the course of decay. As the decay of the atomic state of a given
symmetry $\nu$ involves only photons of the same symmetry, the problem
is similar to the usual one-atom decay problem and here 
we write down the calculated decay constants (referring to Appendix~A1 for
technical details). The decay constant for the symmetric singly excited state is
\begin{figure}[t]
\centering
\includegraphics[width=0.5\textwidth]{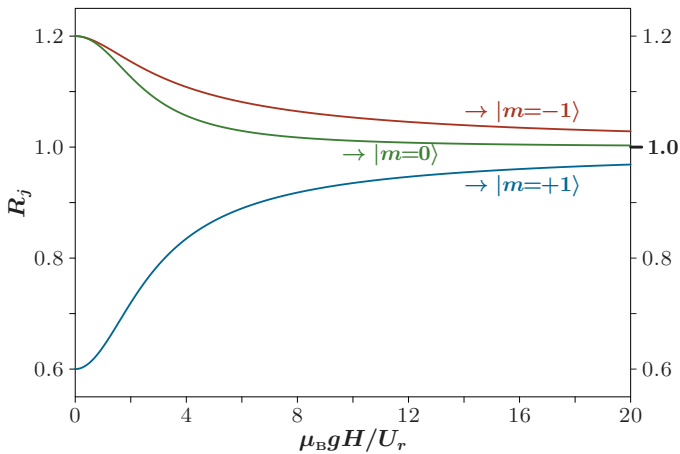}
\caption{Magnetic field dependence of the decay rate coefficients $R_j$ defined by 
Eqs.~(\ref{gamma-a})--(\ref{R-j}). The field ${\bf{H}}$ is directed at the 
angle $\alpha=\alpha_{\textrm{mag}}$ relative to the vector separation ${\bf{r}}$
between the atoms. Arrows indicate to which $|m\rangle$-state 
a given $j$-eigenstate transforms as ${\bf{H}}$ increases.}  
\label{R-j-fig}
\end{figure}
\begin{eqnarray}\label{gamma-s}
\varGamma_{s} \approx 2\varGamma \, ,
\end{eqnarray}
where $\varGamma$ is the decay rate for a single atom
\begin{eqnarray}\label{gamma}
\varGamma= \frac{4}{3}\frac{\omega^3_{eg} d^2_0}{\hbar c^3} \;.
\end{eqnarray}
We see that decay rate of the symmetric states demonstrates the expected
superradiance property and is almost non-sensitive to the magnetic field.
On the contrary, the decay rate of antisymmetric states $\varGamma_a$ is strongly suppressed (subradiant). In the case of a strong magnetic field ($\varOmega_H \gg U_r$) and $\alpha=\alpha_{\textrm{mag}}$, the decay rate of any singly excited 
antisymmetric state is given by
\begin{eqnarray}\label{gamma-a-H}
\varGamma_a\approx{\frac{1}{6}} (k_0r)^2\varGamma \; 
\end{eqnarray}
(see Ref. \cite{MAK} and Sec.~1 of Appendix A).
For the intermediate field magnitudes the decay rate of a state $|\psi^{a}_j\rangle$
takes the form:
\begin{eqnarray}\label{gamma-a}
\varGamma_{aj}=R_{j}\varGamma_a
\; ,
\end{eqnarray}
where the numerical coefficient $R_{j}$ is a combination
\begin{eqnarray}\label{R-j}
R_{j} = 1-\frac{1}{5}\sum\nolimits_{m,m'\neq m}C^{a}_{m j}
(C^{a}_{m'j})^*
\; ,
\end{eqnarray}
of the elements of the matrix $C^{a}_{mj}$ [see Eq.~(\ref{C-matrix})]. With the increase
of the magnetic field the matrix elements $C^{a}_{m j} \rightarrow \delta_{m j}$, and
we return to the expression (\ref{gamma-a-H}).
The dependences of the coefficients $R_{j}$ on the magnetic 
field are shown in  Fig.~\ref{R-j-fig}.

In the general case where the angle $\alpha$ is arbitrary, the dependences 
like shown in Fig.~\ref{R-j-fig} can be obtained using 
Eqs.~(\ref{G-a-cal})--(\ref{G-a-cal-general}) from Appendix A. 
In the explicit form they are
\begin{multline}
R_j={\frac{3}{10}}\left[\left(Q_{j,-1}^a\right)^2+\left(Q_{j,+1}^a\right)^2\right]
\left(3+\cos^2\alpha\right)\\+
{\frac{3}{5}}\left(Q_{j,0}^a\right)^2\left(2-\cos^2\alpha\right)
-{\frac{3}{5}}Q_{j,-1}^aQ_{j,+1}^a\sin^2\alpha\\-
{\frac{3{\sqrt{2}}}{10}}\left(Q_{j,-1}^a+Q_{j,+1}^a\right)Q_{j,0}^a\sin2\alpha\;,
\end{multline}
where $Q_{jm}^a$ are components of the $j$-th antisymmetric eigenvector in the 
$|m\rangle$-basis defined by Eqs.~(\ref{Eig_basis})--(\ref{norma}). 
 
\subsection{Fluorescence spectrum of a doubly excited system}\label{Decay-doubly}
\subsubsection{Time evolution of a doubly excited system}\label{fluo-2}
Now we consider the decay of a doubly excited state of the two-atom system.
For definiteness, we choose the state $|0, 0\rangle$ (i.e., $m_1=m_2=0$)
that can be prepared by 
a laser $\pi$-pulse linearly 
polarized along $z$-direction.
We are searching for the time dependent state of the atom-photon system in the form:
\begin{widetext}
\begin{multline}\label{Pci-2-t}
|\varPsi(t)\rangle = A(t)|0,0\rangle\otimes|{\textsl{vac}}\rangle +
 \sum\nolimits_{{\bf{k}},\lambda,j,\nu}'B^{(\nu)}_{{\bf{k}}\lambda j}(t)
|\psi^{\nu}_j\rangle\otimes b^{\dagger}_{\lambda \nu}({\bf{k}})|{\textsl{vac}}\rangle\\
+ \sum\nolimits_{{\bf{k}}_{1(2)},\lambda_{1(2)},\nu}'
C^{(\nu)}_{{\bf{k}}_1,\lambda_1,{\bf{k}}_2,\lambda_2}(t)|g,g\rangle \otimes 
b^{\dagger}_{\lambda_1 \nu}({\bf{k}}_1)\rangle
b^{\dagger}_{\lambda_2 \nu}({\bf{k}}_2)|{\textsl{vac}}\rangle
\;,
\end{multline}
\end{widetext}
where the function $C^{(\nu)}_{{\bf{k}}_1\lambda_1{\bf{k}}_2\lambda_2}(t)$
is symmetric with respect to the permutation $({\bf{k}}_1,\lambda_1) 
\leftrightarrow ({\bf{k}}_2,\lambda_2)$.
The initial state $|\varPsi(t=0)\rangle=|0,0\rangle\otimes|{\textsl{vac}}\rangle$ 
corresponds to the initial condition $A(t=0)=1$, while all the other 
probability amplitudes are zero.

The emission spectrum in the direction ${\bf{\widehat{k}}}\equiv{\bf{k}}/|{\bf{k}}|$ is 
defined as
\begin{eqnarray}\label{spectrum-def}
\mathcal{S}(\omega, \widehat{{\bf{k}}}) = \sum_{\lambda} \langle \varPsi(t)| 
a^{\dagger}_{{\bf{k}} \lambda}
a_{{\bf{k}} \lambda}|\varPsi(t)\rangle_{t\rightarrow \infty}\;,
\end{eqnarray}
where $k=\omega/c$.
In the limit $t\rightarrow \infty$, only the amplitudes 
$C^{(\nu)}_{{\bf{k}}_1\lambda_1{\bf{k}}_2\lambda_2}(t)$
 in Eq.~(\ref{Pci-2-t})
differ from zero. Accounting for Eqs. (\ref{b1s})--(\ref{b2a}), we find
\begin{eqnarray}\label{spectrum-via-C}
\mathcal{S}(\omega, \widehat{{\bf{k}}} ) = 
2\sum\nolimits_{\lambda,\nu,{\bf{k}}',\lambda'}'
|C^{(\nu)}_{{\bf{k}}\lambda{\bf{k}}'\lambda'}(t\rightarrow\infty)|^2\;.
\end{eqnarray}

\subsubsection{Probability of reaching an antisymmetric state}\label{Reach-antisymm}
The system of Schrodinger equations for the time evolution of the doubly excited
state (\ref{Pci-2-t}) is rather complicated (see 
Eq.~(\ref{dABCdt}) in Appendix B). However, its study can
be simplified due to the great difference of the relevant time scales:
the decay rate $\varGamma_{s}$ through the symmetric channel is much greater
than that $\varGamma_{a}$ through the antisymmetric one.
This means that the time evolution of the amplitude $A(t)={\bar{A}}(t)e^{-2\omega_{eg}t}$ 
of the doubly excited state
is fast and governed mostly by the symmetric channel. Neglecting the influence of
the antisymmetric channel during the short time interval 
$1/\varGamma_{s} \ll t \ll 1/\varGamma_{a}$,
we find the Laplace image defined analogously to Eq.~(\ref{Laplace})
${\bar{A}}[z]$ and the time dependence ${\bar{A}}(t)$:
\begin{eqnarray}\label{laplA0}
\bar{A}[z] \approx \frac{1}{z+\varGamma_{s}/2}\;, \,\,\;\;\; {\bar{A}}(t) = e^{-\varGamma t}
\end{eqnarray}
with $\varGamma_{s} = 2\varGamma$ [see Eq.~(\ref{gamma-s})]. However, during the considered 
short time
interval $1/\varGamma_{s} \ll t \ll 1/\varGamma_{a}$ there is a small but finite
probability of emitting a single ``antisymmetric'' photon with transition
to singly excited atomic states
(with $\nu =a$) described by the second term in equation (\ref{Pci-2-t}).
If such a transition occurs, the further (slow) time evolution of the
system follows the antisymmetric channel. The antisymmetric atomic
state, formed at the short-time interval may be considered as an initial
state for the further evolution. Our current task is to estimate the
probability of the formation of such an antisymmetric state.
This probability is given by
\begin{eqnarray}\label{Prob-definition}
P = \sideset{}{'}\sum_{{\bf{k}},\lambda,j}|B_{{\bf{k}}\lambda j}^a(t)|^2 \, ,
\end{eqnarray}
that actually does not depend on time if it lies in the interval
$1/\varGamma_{s} \ll t \ll 1/\varGamma_{a}$.
To find this quantity it is sufficient to solve Eq.~(\ref{Schr-B-a})
for the amplitudes $B_{{\bf{k}}\lambda j}^a$ neglecting the slow processes
of further decay (i.e. terms with amplitudes $C$) and taking the amplitude $A$
in the form (\ref{laplA0}). As a result we obtain:
\begin{equation}\label{B-a-in}
B_{{\bf{k}}\lambda j}^a\left({\frac{1}{\varGamma_s}} \ll t \ll {\frac{1}{\varGamma_a}}\right)=
-{\frac{2i\left(g_{{\bf{k}}\lambda}^{m=0,g}\right)^{\!*}
\sin\left({\frac{{\bf{kr}}}{2}}\right)C_{0j}^a
}
{E_j^a + \omega_k - 2\omega_{eg} + i\varGamma}}
\, .
\end{equation}
Using Eq.~(\ref{B-a-in}) we find (see derivation from Eq.~(\ref{P-j-derivation}) to 
Eq.~(\ref{P-intermediate}) in Appendix~A) the 
probability
\begin{eqnarray}\label{Prob}
P \approx{\frac{1}{12}} {|k_0 r|^2}.
\end{eqnarray}
of the transition of the doubly excited atomic system to a singly excited antisymmetric state.
This probability has been obtained for the simplified case 
$|k_0r|\ll1$,
where $P$ is small. However, no matter how small it is, its  non-zero value results
in the delayed (at times
 $t\gtrsim 1/\varGamma_a \gg 1/\varGamma_s $) fluorescence
from antisymmetric states. Measuring the ratio of 
quantum yields of the fast
and the delayed fluorescence, one can obtain the quantity
(\ref{Prob}) and, thus, the desired distance $r$ between the atoms.
\subsubsection{Fluorescence spectrum in a strong magnetic field}\label{angular-resolution}
\begin{figure}[t]
\centering
\includegraphics[width=0.5\textwidth]{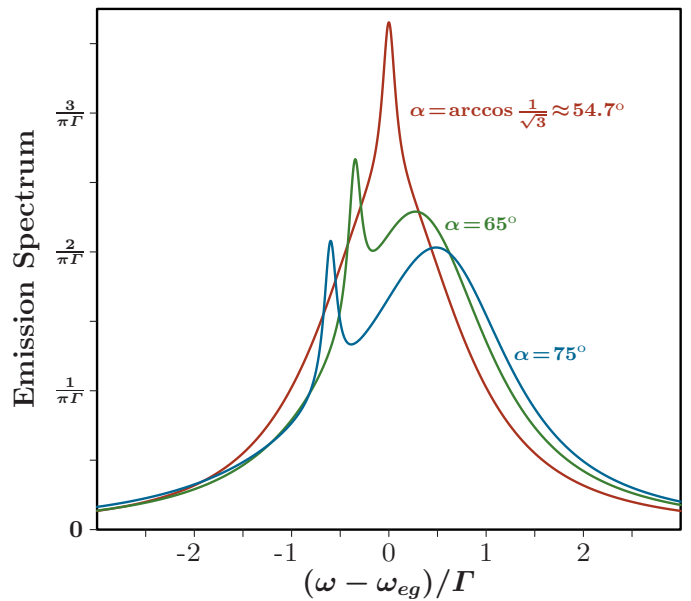}
\caption{Emission spectra of a two-atom system in
a strong magnetic field ${\textbf{H}}$ at different values of the angle $\alpha$ at 
$k_0r=1$. All other preconditions are the same as in Fig.~\ref{spectrum-strong-H}.}  
\label{spectrum-strong-H-1}
\end{figure}

We begin the analysis from a simple but instructive case of a strong magnetic field
($\varOmega_H \gg U_r$), when the effect of the dipole-dipole mixing of states with 
different $m$ is negligible. In this case, the spontaneous decay of the doubly excited 
atomic state $|0,0\rangle$ goes via the singly excited states with $m=0$, either symmetric 
$|s_{0}\rangle$ or antisymmetric $|a_{0}\rangle$ (see Fig. \ref{Fig-decay}).
The formula for the emission spectra defined by 
Eqs. (\ref{Pci-2-t})--(\ref{spectrum-via-C}) can be obtained with just
a little supplement to the method of Ref.~\cite{MAK_YUD} where the general formula for the 
spectrum of cascade spontaneous emission has been derived, including the case of close 
transition frequencies. The only difference is that now there are {\em{two ways}} of
two-photon emission whereas the formula of Ref.~\cite{MAK_YUD} refers to the case of 
a {\em{single way}}. Its modification to our case is derived in Appendix B, and the final 
result is described in terms of the functions
\begin{equation}\label{Lorentz}
\begin{split}
{\mathcal{L}}(\omega;p;q)&={\frac{1}{2\pi}}{\frac{p}{(\omega-q)^2+{\frac{1}{4}}p^2}}\;,\\
{\mathcal{M}}(\omega;p;q)&={\frac{1}{2\pi}}{\frac{\omega-q}{(\omega-q)^2+{\frac{1}{4}}p^2}}
\end{split}
\end{equation}
that are the real and imaginary parts of the complex Lorentzian. 
So, the emission spectrum 
\begin{equation}\label{Somega}
{\mathcal{S}}(\omega)=\int\limits_{{\widehat{{\bf{k}}}}}S(\omega,{\widehat{{\bf{k}}}})\,
d{\widehat{{\bf{k}}}}
\end{equation} 
with $S(\omega,{\widehat{{\bf{k}}}})$ given by 
Eq.~(\ref{spectrum-via-C}) is represented, after the integration over all directions of the 
wave vector, as a sum of two constituents
\begin{equation}\label{Ss+Sa}
{\mathcal{S}}(\omega)={\frac{\varGamma_s}{\varGamma_s+\varGamma_a}}{\mathcal{P}}_s(\omega)+
{\frac{\varGamma_a}{\varGamma_s+\varGamma_a}}{\mathcal{P}}_a(\omega)\;,
\end{equation}
they being expressed as combinations of the functions (\ref{Lorentz}) as
\begin{multline}{\label{PsPa}}
{\mathcal{P}}_s(\omega)=
\left(1+{\frac{
2\varGamma_s(\varGamma_s+\varGamma_a)}{4\varDelta^2+(\varGamma_s+\varGamma_a)^2}}\right)
{\mathcal{L}}(\omega;\varGamma_s;\omega_{eg}+\varDelta/2)\\
\left.+\left(1-{\frac{
2\varGamma_s(\varGamma_s+\varGamma_a)}{4\varDelta^2+(\varGamma_s+\varGamma_a)^2}}\right)
{\mathcal{L}}(\omega;2\varGamma_s+\varGamma_a;\omega_{eg}-\varDelta/2)\right.\\
\left.-{\frac{4\varGamma_s\varDelta}{4\varDelta^2+(\varGamma_s+\varGamma_a)^2}}\right.\\
\times\big[
{\mathcal{M}}(\omega;\varGamma_s;\omega_{eg}+\varDelta/2)
-{\mathcal{M}}(\omega;2\varGamma_s+\varGamma_a;\omega_{eg}-\varDelta/2)\big]\;,\\
{\mathcal{P}}_a(\omega)=\left(1+{\frac{
2\varGamma_a(\varGamma_s+\varGamma_a)}{4\varDelta^2+(\varGamma_s+\varGamma_a)^2}}\right)
{\mathcal{L}}(\omega;\varGamma_a;\omega_{eg}-\varDelta/2)\\
\left.+\left(1-{\frac{
2\varGamma_a(\varGamma_s+\varGamma_a)}{4\varDelta^2+(\varGamma_s+\varGamma_a)^2}}\right)
{\mathcal{L}}(\omega;\varGamma_s+2\varGamma_a;\omega_{eg}+\varDelta/2)\right.\\
\left.-{\frac{4\varGamma_a\varDelta}{4\varDelta^2+(\varGamma_s+\varGamma_a)^2}}\right.\\
\times\big[
{\mathcal{M}}(\omega;\varGamma_a;\omega_{eg}-\varDelta/2)
-{\mathcal{M}}(\omega;\varGamma_s+2\varGamma_a;\omega_{eg}+\varDelta/2)\big]\;.
\end{multline}
Here $\varDelta$ means the splitting between the singly excited states $s_0$ and $a_0$.
Examples calculated using Eqs.~(\ref{Lorentz})--(\ref{PsPa}) are shown in 
Fig.~\ref{spectrum-strong-H}. The decay via the symmetric channel is fast and leads to a relatively broad spectral contour. A considerably less probable decay via the antisymmetric channel is slow and leads to a weak but very narrow spectral peak
at the background of a broad contour of the symmetric decay.
The difference $\varDelta\omega$ between 
the centers of the broad and narrow peaks
is determined by Eqs.~(\ref{High-H-m0}) and (\ref{High_field non-mag}) for $\nu = 1$
($s$-channel) and $\nu = -1$ ($a$-channel):
it is of the order of the dipole-dipole interaction, i.e.
\begin{eqnarray}\label{Delta non-mag}
\varDelta\omega = E^s_{m=0} - E^a_{m=0} = 2(1- 3\cos^2{\alpha})U_r 
\end{eqnarray}
when the angle $\alpha$ is not close to the magic one $\alpha_{\textrm{mag}}$, and it
is almost zero when $\alpha \approx \alpha_{\textrm{mag}}$ [see Eq.~(\ref{High-H-m0})].
With deflection of the angle $\alpha$ from $\alpha_{\textrm{mag}}$, the divergence of the 
two peaks 
rapidly slows down as the distance between the atoms grows. This is 
due to the $r^{-3}$ dependence of $U_r$ on $r$. One illustration 
is given in Fig.~\ref{spectrum-strong-H-1} where $k_0r=1$, whereas 
$k_0r=0.5$ in Fig.~\ref{spectrum-strong-H}.
The difference of two pictures can be seen from the angle values attached to the curves: 
they are close to $\alpha_{\textrm{mag}}$ in Fig.~\ref{spectrum-strong-H} while they  
considerably deviate from $\alpha_{\textrm{mag}}$ in Fig.~\ref{spectrum-strong-H-1}. 

For the completeness, below we consider also the spectrum of the delayed
fluorescence, measured at times much longer than the decay time $1/\Gamma_s$ through
the symmetric channel.

\subsubsection{Spectrum of the delayed decay of the doubly excited system}

Actually the delayed decay of the doubly excited two-atom system is a decay
of a singly excited metastable state where the doubly excited system falls to
(with a small probability [\ref{Prob})] after emission of a antisymmetric photon
at an instant $t_{\textrm{in}}$ within a short time interval $~1/\Gamma_s$.
This antisymmetric state is a superposition containing all possible
polarizations and wave vectors of the emitted photon:
\begin{eqnarray}\label{Psi-2-in}
|\varPsi(t_{\textrm{in}})\rangle =\sideset{}{'} \sum_{{\bf{k}},\lambda,j}
B^{(a)}_{{\bf{k}}\lambda j}
|\psi^{a}_j\rangle\otimes b^{\dag}_{\lambda a}({\bf{k}})|{\textsl{vac}}\rangle \, ,
\end{eqnarray}
where the amplitudes $B^{(a)}_{{\bf{k}}\lambda j}$ are given by Eq.~(\ref{B-a-in}).
The state (\ref{Psi-2-in}) may be considered as an ``initial state''
($t_{\textrm{in}}\approx 0$) of further time evolution. This evolution may, in principle,
be described by the wave function (\ref{Pci-2-t}) with only antisymmetric components and 
with $A(t > t_{\textrm{in}})=0$.
However, such a representation is not convenient for calculation of the
delayed spectrum, because the state (\ref{Psi-2-in}) keeps information about
the primary photon emitted at the time $t_{\textrm{in}}$, while the delayed spectrum
is measured at a considerably later time, when the primary photon is already far from the 
registration region. It is difficult to distinguish between 
the the primary and secondary photons in the framework of Schr\"{o}dinger
state evolution. To overcome this problem we develop the density matrix
approach which eliminates excessive information about the primary photon.

The initial density matrix operator $\hat{\rho}_{\textrm{in}}$ of the atomic subsystem
is determined by the state (\ref{Psi-2-in}):
\begin{multline}\label{rho-in}
 \hat{\rho}_{\textrm{in}} = {\mathrm{Tr}}_{\textrm{ph}}\left\{|\varPsi(t_{\textrm{in}})\rangle \langle\varPsi(t_{\textrm{in}})|\right\}\\
 =\sideset{}{'}\sum_{{\bf{k}},\lambda,j,j'}B^{(a)}_{{\bf{k}}\lambda j}
 B^{(a)}_{{\bf{k}}\lambda j'}|\psi^{a}_j\rangle \langle\psi^{a}_{j'}|
 \; ,
\end{multline}
where the trace is taken over the photon degrees of freedom.
This operation allows one to forget about a primary photon
emitted at the early time $t_{\textrm{in}}$, but the price paid is that
the matrix $\hat{\rho}_{\textrm{in}}$ corresponds not to a pure but
to a mixed state. The initial density matrix $\widehat{R}_{\textrm{in}}$
of the whole atomic-photon system is constructed as a direct
product of the atomic density matrix $\hat{\rho}_{\textrm{in}}$
and the density matrix $|{\textsl{vac}}\rangle \langle {\textsl{vac}}| $ of photon vacuum:
\begin{eqnarray}\label{R-in}
\widehat{R}_{\textrm{in}} = \hat{\rho}_{\textrm{in}}\otimes |{\textsl{vac}}\rangle
\langle{\textsl{vac}}|\;.
\end{eqnarray}
The further temporal dynamics of the system is governed by the Liouville equation:
\begin{eqnarray}\label{Lioville}
&&i\frac{d}{dt}\widehat{R}(t)=[\widehat{\mathcal{H}}_{a},\widehat{R}]\, .
\end{eqnarray}
for the system density matrix $\widehat{R}(t)$ with the initial condition
$\widehat{R}(t=0)=\widehat{R}_{\textrm{in}}$.
The Hamiltonian $\widehat{\mathcal{H}}_{a}$ in (\ref{Lioville}) is the sum
of the atom Hamiltonian $\widehat{H}^{(1)}_a$ (\ref{Matrix}),
the $a$-part of the photon Hamiltonian (\ref{H-photon}), and the
$a$-part of the interaction Hamiltonian (\ref{H-int-1-b}):
\begin{multline}\label{Ham-spec}
\widehat{\mathcal{H}}_{a} = \sideset{}{'}\sum_{{\bf{k}},\lambda}
\hbar\omega(k)b_{\lambda a}^{\dag}({\bf{k}})b_{\lambda a}({\bf{k}})
+\sum_j E_j|\psi_j^a\rangle\langle\psi_j^a|\\
+\sideset{}{'}\sum_{{\bf{k}},\lambda,j}[F_{{\bf{k}}\lambda}^j|\psi_j^a
\rangle \langle g,g|b_{\lambda a}({\bf{k}}) + \mathrm{H.c.}
 ]\; ,
\end{multline}
where
\begin{eqnarray}\label{F}
F_{{\bf{k}}\lambda}^j=2i\sum_{m} g_{{\bf{k}}\lambda}^m
\sin\left({\frac{{\bf{kr}}}{2}}\right)C_{mj}^a \, .
\end{eqnarray}
It should be noted that the developed formalism describes a
``conditional'' evolution of the system---under the condition
that the decay goes through the antisymmetric channel. This
is why the traces of the initial density matrices $\hat{\rho}_{\textrm{in}}$
and $\widehat{R}_{\textrm{in}}$ equal not to unity but to the 
probability
(\ref{Prob}) of falling to the antisymmetric channel.

In terms of the density matrix ${\widehat{R}}(t)$, the spectrum of
the delayed fluorescence in the direction $\widehat{{\bf{k}}}$ is given by
\begin{equation}
S(\omega,\widehat{{\bf{k}}})\simeq \sum_{\lambda}
\mathrm{Tr}\left\{
b_{\lambda a}^{\dag}({\bf{k}}) b_{\lambda a}({\bf{k}})\widehat{R}(t)
\right\}_{t \rightarrow \infty;\,k = \omega/c} \, ,
\end{equation}
where it is assumed that the measurement is performed during
the time interval
$(t_1, t_2)$ with $1/\Gamma_s \ll t_1 \ll 1/\Gamma_a \ll t_2$.

The matrix $\widehat{R}$ can be represented in the form
\begin{multline}\label{R-state}
\widehat{R}(t)=\sum_{j,j'}\mathcal{A}_{jj'}(t)|\psi_j^a\rangle\langle\psi_{j'}^a|\otimes|{\textsl{vac}}\rangle\langle {\textsl{vac}}|\\
+\sideset{}{'}\sum_{{\bf{k}},\lambda,j}[ \mathcal{B}^{j}_{{\bf{k}} \lambda}(t)|g,g\rangle
\langle\psi_{j}^a|\otimes|{\bf{k}}\lambda\rangle\langle {\textsl{vac}}|
+ \mathrm{H.c.}]
\\
+\sideset{}{'}\sum_{{\bf{k}},\lambda,{\bf{k}}',\lambda'}
\mathcal{D}_{{\bf{k}}\lambda;{\bf{k}}'\!\lambda'}(t)
|g,g \rangle\langle g,g|\otimes|{\bf{k}}\lambda\rangle\langle {\bf{k}}'\!\lambda'|\; ,
\end{multline}
where $|{\bf{k}}\lambda\rangle = b_{\lambda a}({\bf{k}})|{\textsl{vac}}\rangle$. The initial condition
$\widehat{R}(t=0)=\widehat{R}_{\textrm{in}}$ takes the form: $\mathcal{A}_{jj'}(0) = 
(\hat{\rho}_{\textrm{in}})_{jj'}$,
$\mathcal{B}^{j}_{{\bf{k}} \lambda}(0)=0$, 
$\mathcal{D}_{{\bf{k}}\lambda;{\bf{k}}'\!\lambda'}(0) = 0$.
In this representation, the spectrum has the form:
\begin{eqnarray}\label{spectrum-delayed}
S(\omega, \widehat{\bf{k}}) = \sum_{\lambda}\mathcal{D}_{{\bf{k}}\lambda; {\bf{k}} \lambda}(t\rightarrow\infty)|_{k = \omega/c}\, .
\end{eqnarray}
The Liouville equation (\ref{Lioville}) leads to a system
of linear differential equations for the amplitudes entering
the representation (\ref{R-state}). 
These equations can be
solved by the Laplace transform, a bit boring but
straightforward calculations. In principle, this allows one
to find arbitrary correlation functions of the emitted light
including the spectrum (\ref{spectrum-delayed}).

\begin{figure}[t]
\centering
\includegraphics[width=0.5\textwidth]{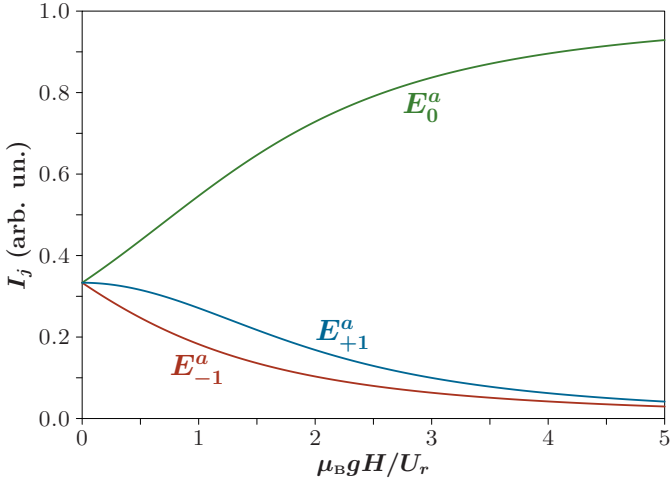}
\caption{Magnetic field dependence of the intensities $I_j$ of peaks of the delayed 
fluorescence (see the text) at the frequencies $E_j^a/\hbar$.  Their ratios are defined by 
Eq.~(\ref{I-j}), and they are normalized so that $\sum_jI_j=1$. The field ${\bf{H}}$ is directed at the 
angle $\alpha=\alpha_{\textrm{mag}}$ relative to the vector separation ${\bf{r}}$
between the atoms. Initially excited state is $|m=0,m=0\rangle$.}  
\label{I-j-fig}
\end{figure}

However, the main features of the delayed spectrum can be
understood based on the presented here scenario
of the decay. Namely, the spectrum consists of three very
narrow peaks of widths $\sim \Gamma_{aj}$ (\ref{gamma-a});
the peaks' centers correspond to energies of three antisymmetric
singly-excited states ($j=-1,0,1$) of the atomic subsystem
in the magnetic field. The peak weight $I_j$ (i.e., the
emission power collected from all the angles and
integrated over frequencies within the peak widths) is
proportional to the initial population $P_j$ of the
state $j$:
$I_j \propto P_j \equiv (\hat{\rho}_{\textrm{in}})_{jj}$.
The latter quantity is calculated with the use of 
Eqs.~(\ref{B-a-in}) (\ref{rho-in}) (see Eq.~(\ref{P-j-intermediate}) in Appendix~A).
In the case of our major interest, when the angle $\alpha$
between the vector connecting the two atoms and the
magnetic field is a magic one, Eq.~(\ref{P-j-intermediate}) takes a simple form:
\begin{eqnarray}\label{Prob-j}
P_j = {\frac{1}{12}}|k_0 r|^2|C^{a}_{mj}|^2 \, ,
\end{eqnarray}
where the index $m$ characterizes the original doubly excited state
 $|m,m \rangle$, and
the matrix $\widehat{C}^{a}$ is determined by Eqs.~(\ref{C-matrix}), (\ref{Eig_basis}), and (\ref{norma}). Here we give ratios of the peak weights for the considered decay of the doubly excited state $|0,0\rangle$ prepared by a linearly
polarized $\pi$-pulse:
\begin{multline}\label{I-j}
 I_{-1} : I_0 : I_1 \\
=\left[ \frac{q_{-1}(E^a_{-1}+U_r)}
{\varOmega_H + E^a_{-1}+U_r }\right]^2 : q^2_0 :
\left[\frac{q_{+1}(E^a_{1}+U_r)}
{\varOmega_H - E^a_{1}-U_r }\right]^2 \; ,
\end{multline}
where $E^a_{j}$ and $q_j$ are determined by Eqs.~(\ref{SA}) and (\ref{norma}),
respectively. An example of the dependence of peak weights on the magnitude of the magnetic 
field is shown in Fig.~\ref{I-j-fig}.

Using the developed formalism one can also calculate the integral intensities
$I_j({\widehat{{\bf{k}}}})$ of the lines at the frequencies $E_j^a/\hbar$ 
for an arbitrary direction ${\widehat{{\bf{k}}}}$ of observation.
These quantities 
and the relation between them turn out to depend on ${\widehat{{\bf{k}}}}$.

\section{Determination of the mutual orientation of the two atoms}

\begin{figure}[t]
\centering
\includegraphics[width=1\textwidth]{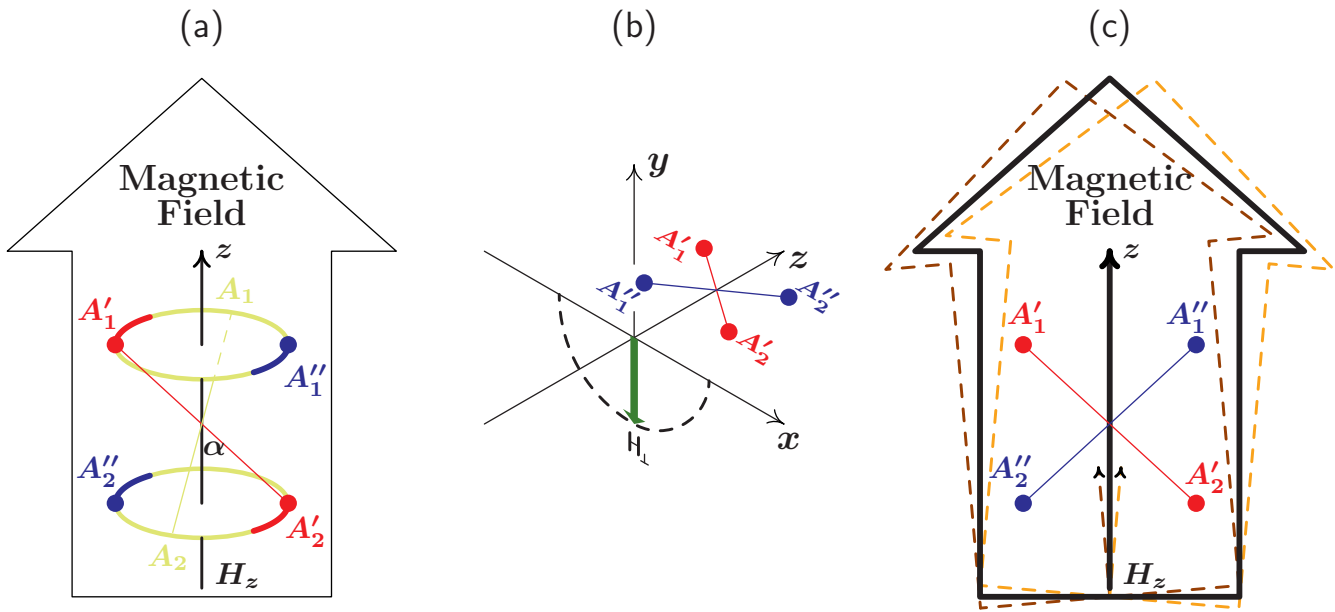}
\caption{Determination of the direction of the vector separation ${\bf{r}}$ using 
subsequently three configurations of the magnetic fields. 
(a) Superimposition of the narrow and broad peaks of the spectrum in 
Fig.~\ref{spectrum-strong-H} by varying of the ${\textbf{H}}$ direction 
leads to determination of the cone of possible directions of ${\bf{r}}$ and fixes the axis $z$. 
(b) The rotation of ${\bf{H}}_\perp$ in the plane perpendicular to $z$ allows one to fix 
the axises $x$ and $y$ when the separation of the two peaks is maximal, i.e., the angle between 
${\bf{H}}_\perp$ and ${\bf{r}}$ equals to $\pi/2$.
(c) The choice between the two remaining possible directions can be made by slight 
deflection
of the ${\bf{H}}$ direction from $z$---as seen from Fig.~\ref{spectrum-strong-H} the counter
clockwise deflection leads to the red shift of the narrow peak, and vice versa.}
\label{Rotation}
\end{figure}

Sensitivity of the 
fluorescence spectrum of the doubly 
excited system to the angle 
$\alpha$ between the vector separation ${\bf{r}}$ and the 
magnetic field ${\bf{H}}$
(see Sec.~\ref{angular-resolution} and
 Fig~\ref{spectrum-strong-H}) allows one to find the direction
of ${\bf{r}}$ from an experiment. As follows from Eq.~(\ref{Delta non-mag}) the 
positions of the broad and narrow central peaks almost coincide when 
$\alpha \approx \alpha_{mag}$.
Thus, rotating the direction of the magnetic field (while keeping the propagation direction and polarization of the pumping light orthogonal and parallel to the magnetic field, respectively) and measuring the emission spectrum, one
can find the configuration where the 
positions of two peaks in
Fig.~\ref{spectrum-strong-H} (or in Fig.~\ref{spectrum-strong-H-1}) coincide. This means
that the angle between the found direction ${\mathbf{\widehat{h}}}_1$ of the
magnetic field ${\bf{H}}$ and the separation vector ${\bf{r}}$ equals to the magic
angle $\alpha_{\textrm{mag}}$. However, this condition determines not the direction 
$\widehat{\mathbf{r}}$ of $\bf{r}$ but the whole cone of possible directions
(see Fig.~\ref{Rotation}a). To find $\widehat{\mathbf{r}}$ one should repeat the routine 
searching for some other suitable direction $\widehat{\mathbf{h}}_2$ of the magnetic field 
such that the two spectral peaks coincide, so that 
${\widehat{\mathbf{r}}}{\widehat{\mathbf{h}}}_2 =
\cos\alpha_{\textrm{mag}}$. From these measurements and the elementary geometrical analysis
presented in Appendix C,
one finds the desired vector $\widehat{\mathbf{r}}$ between the atoms in the form:
\begin{multline}\label{l-direction}
\widehat{\mathbf{r}} = \frac{\cos{(\alpha_{\textrm{mag}})}}{1+\cos{\vartheta_{12}}}
[\widehat{\mathbf{h}}_1 + \widehat{\mathbf{h}}_2 ] \\
\pm\frac{1}{\sin{\vartheta_{12}}}\sqrt{1 - \frac{2\cos^2{(\alpha_{\textrm{mag}})}}{1+
\cos{\vartheta_{12}}}}
[\widehat{\mathbf{h}}_1 \times \widehat{\mathbf{h}}_2 ]
 \, ,
\end{multline}
where $\vartheta_{12}$ is the angle between 
$\widehat{\mathbf{h}}_1$ and $\widehat{\mathbf{h}}_2$.
The remaining uncertainty 
[due to the $\pm$ sign in Eq.~(\ref{l-direction})]
can be eliminated by repeating the routine and finding the third suitable
direction $\widehat{\mathbf{h}}_3$ of the magnetic field.

The described procedure presents an example of determining the direction of the vector separation. Another possibility is shown in Fig. \ref{Rotation}.

As described in Sec.~\ref{Reach-antisymm}, the information 
about the length of 
$\bf{r}$ can be extracted by measuring the ratio of the quantum yields 
of the fast and the delayed fluorescence and using Eq.~(\ref{Prob}).
 Together with the information about the direction of $\bf{r}$ 
[Eq.~(\ref{l-direction})] 
this provides full knowledge about 
the relative arrangement of two atoms.

\onecolumngrid

\newpage
\twocolumngrid

\section{Conclusion}
We have described fluorescence of an excited system of two
closely located identical atoms, each atom
having a non-degenerate ground state $|g\rangle$ of the angular
momentum $J=0$ and a triply degenerate excited state
$|J=1, m =0, \pm 1 \rangle$. We accounted for the resonant
dipole-dipole interaction between close atoms and its competition
with the Zeeman splitting of the levels in an external magnetic field.
The fluorescence spectrum and its time behavior depend on the
strength and direction of the magnetic field.
The emission of a doubly excited system possesses two
very different time regimes: for the ensemble, fast strong emission
pulse (of the duration $\sim 1/\varGamma_s$) is followed
by a weak but long (of the duration $\sim 1/\varGamma_a$)
fluorescence from metastable states where the
system can fall with a small but finite probability
after the excitation.
A very sensitive dependence of the fluorescence spectrum
on the direction of the magnetic field allows one to
determine the direction of the vector connecting
the two atoms. On the other hand, measuring the ratio of the
fast and the delayed emission powers one can extract
the information about the distance between the two atoms.
Thus, one can get a complete information about the relative
position of two atoms at nanoscale by entirely optical means.

We have studied the simplest realistic two-particle model. A similar analysis  
may be applied to other systems: quantum dots, color centers, molecules in matrices, etc.

\acknowledgements
The work was partially supported by Russian Foundation
of Basic Research (grant No. 15-02-05657a) and by Basic research program of HSE.

\section*{Appendix A: Details on interaction with the electromagnetic field}
\renewcommand{\theequation}{A\arabic{equation}}
\setcounter{equation}{0}
We use the standard representation of the electric field operator in terms
of the photon creation and annihilation operators
$a^{\dag}_{{\bf k}, \lambda}$ and $a_{{\bf k}, \lambda}$:
\begin{equation}\label{E}
{\widehat{{\bf{E}}}}({\bf{r}}) =
\sum_{{\bf{k}}, \lambda}\sqrt{\frac{2\pi\hbar \omega_{k}}{V}}
\left[ e^{i{\bf{kr}}} {\bf{e}}_{{\bf{k}}\lambda}a_{{\bf{k}} \lambda}
+ \mathrm{H.c.} \right]
\; ,
\end{equation}
where $V$ is the quantization volume, ${\bf{e}}_{{\bf{k}}\lambda}$ is
the polarization vector ($\lambda =1,\, 2$) of the photon of a wave
vector ${\bf{k}}$ and frequency $\omega_{k}$.
In the dipole approximation, the
atom-field interaction operator is $-{\widehat{{\bf{d}}}}\!\!\cdot\!\!{\widehat{{\bf{E}}}}$.
Its matrix element corresponding in RWA to the atom 
transition $|g\rangle\leftrightarrow|m\rangle$ ($g$ is the ground state, and 
$|m\rangle$ is one of the excited states $|e^{(m)}\rangle$ with $m=0, \pm 1$)  
is given by
\begin{equation}\label{g-app}
g_{{\bf{k}} \lambda}^{m,g}= \sqrt{\frac{2\pi\hbar \omega_{k}}{V}}
\langle m|{\widehat{{\bf{d}}}}|g\rangle{\bf{e}}_{{\bf{k}} \lambda} \; .
\end{equation}
Using RWA, we can
replace the photon frequency $\omega_{k}$ in Eq.~(\ref{g-app}) by
the resonance transition frequency $\omega_{eg}$. In the considered
model of atom levels (of the angular momentum $J=0$ in the ground state,
and $J=1$ in the excited ones), the corresponding wave
functions have a structure typical for a particle in a centrosymmetric field:
$\Phi_{J=0}({\bf{r}}) = R_{0}(r)$,
$\Phi_{J=1, m=\pm 1}({\bf{r}}) = R_{1}(r)\sin{\theta_{{\bf{r}}}}e^{\pm i
\varphi_{{\bf{r}}}}$,
$\Phi_{J=1, m=0}({\bf{r}}) = \sqrt{2}R_{1}(r)\cos{\theta_{{\bf{r}}}}$, where
$\theta_{{\bf{r}}}$ and $\varphi_{{\bf{r}}}$ are 
the polar and the azimuthal angles
of the radius-vector ${\bf{r}}$, $R_{0}(r)$ and $R_{1}(r)$ are the radial functions
in the ground and excited states, respectively.
Non-zero matrix elements $\langle m|{\widehat{\bf{d}}}|g\rangle$ can be expressed in
terms of a single quantity $d_0\equiv \langle 0|\hat{d}_{z}|g\rangle$,
namely: $\langle \pm 1|\hat{d}_{\pm}|g\rangle = \sqrt{2}d_0$, where
$\hat{d}_{\pm}\equiv \hat{d}_{x}\pm i\hat{d}_{y}$.

For calculations one should specify the choice of the
polarization vectors. Representing Cartesian coordinates of
a wave vector ${\bf{k}}$ in the form
\begin{eqnarray}\label{wave_vector}
{\bf{k}}=|{\bf{k}}|(\sin{\theta}\cos{\varphi}, \sin{\theta}\sin{\varphi}, \cos{\theta}) \; ,
\end{eqnarray}
where $\theta$ and $\varphi$ are the polar and the azimuthal angles in the momentum space,
we choose two vectors of the linear polarization in the following form:
\begin{equation}\label{e1}
{\bf{e}}_{{\bf k}1}= (-\sin{\varphi},\, \cos{\varphi},\, 0) \; , \\
\end{equation}
\begin{equation}
{\bf{e}}_{{\bf {k}}2}= \frac{[{\bf{k}}\times{\bf{e}}_{{\bf{k}}1}]}
{\Big|[{\bf{k}}\times{\bf{e}}_{{\bf{k}}2}]\Big|} =
 (\cos{\varphi}\cos{\theta},\, \sin{\varphi}\cos{\theta},\, -\sin{\theta}) \, . \label{e2}
\end{equation}
It is easy to check that these vectors are perpendicular to each other and to 
${\textbf{k}}$.
Then the matrix elements (\ref{g-app}) take the form:
\begin{equation}
\left\{\begin{array}{c}g_{{\bf{k}} 1}^{\pm 1,g}\\
g_{{\bf{k}} 2}^{\pm 1,g}\end{array}\right\}={\sqrt{{\frac{\pi\hbar\omega_{eg}}{V}}}}d_0
e^{\mp i\varphi}\left\{\begin{array}{c}\mp i\\ \cos\theta\end{array}\right\}\;,
\end{equation}
\begin{equation}\label{g-0}
g_{{\bf{k}} 1}^{0,g} = 0\;,\;\;\;\;
g_{{\bf{k}} 2}^{0,g}= -{\sqrt{{\frac{2\pi\hbar\omega_{eg}}{V}}}}d_0\sin{\theta}\;.
\end{equation}

Replacing the original set of photon operators 
$a_{{\bf{k}}\lambda}$ used in
the Hamiltonians (\ref{H-int-1}) and (\ref{H-int-2}) by the properly symmetrized
operators (\ref{b1s})--(\ref{b2a}),   
one should restrict the space of photon
wave vectors by the upper hemisphere ($k_z \geqslant 0$) to avoid double counting and to
provide the correct hold true the 
commutation relations (\ref{b-c-commutators}).
Correspondingly, in the transformation of Hamiltonians from
(\ref{H-int-1}) and (\ref{H-int-2}) to 
(\ref{H-int-1-b}) and (\ref{H-int-2-b})
we used the symmetry properties of the polarization vectors (\ref{e1}) and
(\ref{e2}): ${\bf{e}}_{-{\bf{k}}, 1} = -{\bf{e}}_{{\bf{k}}, 1}$ and
${\bf{e}}_{-{\bf{k}}, 2} = {\bf{e}}_{{\bf{k}}, 2}$.

For further references we write down a useful relation:
\begin{widetext}
\begin{multline}\label{sum-gg}
\sum_{\lambda}g_{{\bf{k}} \lambda}^{m,g} (g_{{\bf{k}} \lambda}^{m',g})^* =
\frac{\pi\hbar\omega_{eg} d^2_0}{V}G_{mm'} 
\;,\;\;\;\;\;\; 
G_{mm'}=\begin{pmatrix} 1+\cos^2\theta & -\sqrt{2}\sin{\theta}\cos{\theta}e^{i\varphi} & 
-\sin^2{\theta}e^{2i\varphi} \\ -\sqrt{2}\sin{\theta}\cos{\theta}e^{-i\varphi} & 
2\sin^2{\theta} & -\sqrt{2}\sin{\theta}\cos{\theta}e^{i\varphi} \\ 
-\sin^2{\theta}e^{-2i\varphi} & -\sqrt{2}\sin{\theta}\cos{\theta}e^{-i\varphi} & 
1+\cos^2\theta \end{pmatrix}
\end{multline}
\end{widetext}
that directly follows from equations (\ref{e1})--(\ref{g-0}).

\subsection*{1. Decay of a singly excited state of the two-atom system}
\label{A-single}
Consider the two-atom system prepared at the time $t=0$
in one of the eigenstates $\psi^{\nu}_j$ of a definite symmetry $\nu = s,a$.
The time evolution of the atom-photon system is described by
the wave function (\ref{Psi1-t}), that obeys the Schr\"{o}dinger equation
with the Hamiltonian (\ref{H-int-1}). It is convenient to represent
$|\varPsi(t)\rangle$ in the form 
$\exp(-iE^{\nu}_j t)|\widetilde{\varPsi}(t)\rangle$
with slowly varying amplitudes. After the Laplace transformation
\begin{equation}\label{Laplace}
|\varPsi(t)\rangle \rightarrow |\varPsi[z]\rangle = \int_0^{\infty}\, dt\, e^{-zt}|\varPsi(t)\rangle
\; ,
\end{equation}
the Schr\"{o}dinger equation for (\ref{Psi1-t}) can be written as
the system of equations for
the Laplace transforms of the slowly varying amplitudes 
their symbol tilde is omitted for brevity):
\begin{equation}\label{Schr-Lapl-Psi-1}
\begin{split}
& [z+ i(\omega_k-E^{\nu}_j)]B^{\nu}_{{\bf{k}}\lambda}[z] =
 2i A^{\nu}_j[z]f^*_{\nu}({\bf{kr}})\sum_m
 (g^{m,g}_{{\bf{k}}\lambda}C^{\nu}_{mj})^*\;, \\
& zA_j^{\nu}[z] = 1 + 2i\sideset{}{'}\sum_{{\bf{k}},\lambda, m}g^{m,g}_{{\bf{k}}\lambda}
C^{\nu}_{mj}f_{\nu}({\bf{kr}}) B^{\nu}_{{\bf{k}}\lambda}[z]
\; ,
\end{split}
\end{equation}
where $f_{\nu}(\xi)=\cos{\xi/2}$ for $\nu = s$ and $f_{\nu}(\xi)=i\sin{\xi/2}$ for 
$\nu = a$.
It is assumed that the energy differences between the levels $E^{\nu}_j$ of the atomic
system are large as compared with the radiative widths of these levels. This is why the
system (\ref{Schr-Lapl-Psi-1}) ignores mixing of states with different $j$.
Representing 
$B^{\nu}_{\vec{k}\lambda}[z]$ in terms of $A^{\nu}_j[z]$ we obtain the following
equation for 
$A^{\nu}_j[z]$:
\begin{equation}\label{A-nu}
\left[z\!+\!\!\sum_{m,m'}\sideset{}{'}\sum_{{\bf{k}},\lambda}\frac{
4 g^{m,g}_{{\bf{k}}\lambda}C^{\nu}_{mj}
|f_{\nu}({\bf{kr}})|^2
(g^{m',g}_{{\bf{k}}\lambda}C^{\nu}_{m'j})^*}
{z+ i(\omega_{{\bf{k}}}-E^{\nu}_j)} \right]\!A^{\nu}_j[z]\!\!=\!\!1
\; .
\end{equation}
The second term in the square brackets describes both the radiative width and the energy
shift. Neglecting the latter we represent (\ref{A-nu}) as
$[z+ \varGamma_{\nu j}/2]A^{\nu}_j[z]=1$,
where the term $\varGamma_{\nu j}/2$ results from the pole integration over
$\omega_k$ near the resonance $\omega_k \approx E^{\nu}_j$:
\begin{equation}\label{Gamma-integral}
\varGamma^\nu_{j} = 4\omega^3_0 d^2_0\int'\frac{d\hat{o}}{4\pi}
|f_{\nu}({\bf{kr}})|^2
\sum_{m,m'} G_{mm'}C^{\nu}_{mj}C^{{\nu} *}_{m'j}
\; .
\end{equation}
Here Eq.~(\ref{sum-gg}) is used, and the integration is performed over the 
hemisphere in the polar 
angles $\theta$ ($0 \leqslant \theta \leqslant \pi/2$) and $\varphi$ $0\leqslant\varphi
\leqslant2\pi$) that describe directions 
of the photon wave-vector ${\bf{k}}$ while its length is kept constant 
($|{\bf{k}}|=k_0 = \omega_{eg}/c$), and $d{\hat{o}}=\sin\theta\,d\theta\,d\varphi$.

Let us consider first the decay of a symmetric state ($\nu = s$). For closely
located atoms one has
$f_{s}({\bf{kr}}) =\cos({\bf{kr}}/2) \approx 1$. The angular integration in 
(\ref{Gamma-integral})
results in $\int' G_{mm'} d\hat{o}/(4\pi) = (2/3)\delta_{mm'}$, so the 
remaining
summation of $C^s_{mj}C^{s *}_{mj}$ over $m$ gives unity 
due to that 
the matrix $\widehat{C}$ is unitary.
So, we arrive at the 
expression 
$\varGamma^s_{j} = 8\omega^3_0 d^2_0/3\hbar c^3 = 2\varGamma$, 
presented by Eqs. (\ref{gamma-s}) and
(\ref{gamma}) in the main text. Thus, the decay rate of the symmetric states 
is a robust quantity insensitive to the magnetic field.

For antisymmetric states ($\nu = a$) the situation is more complicated.
Using the definition of ${\bf{r}}$ in the beginning of the Sec.~II and also
Eq.~(\ref{wave_vector}), the function $f_{a}({\bf{kr}}/2)$
entering Eq. (\ref{Gamma-integral}) is approximated as
$i k_0 r [\sin{\alpha}\sin{\theta}\cos{\varphi} + \cos{\alpha}\cos{\theta}]/2$.
Then the expression (\ref{Gamma-integral}) for $\varGamma^a_j$ takes the form
\begin{eqnarray}\label{Gamma-a-expr}
\varGamma^a_{j} = \frac{3}{4}\varGamma k^2_0 r^2
\sum_{m,m'}\mathcal{G}_{mm'}C^{a}_{mj}(C^{{a}}_{m'j})^*
\; ,
\end{eqnarray}
where
\begin{equation}\label{G-a-cal}
\mathcal{G}_{mm'} =
\frac{1}{4\pi}\int\limits^{{\frac{\pi}{2}}}_0 \sin{\theta} \int\limits^{2\pi}_0
 \frac{({\bf{kr}})^2}{k^2r^2}G_{mm'}\,d\varphi\,d\theta \; .
\end{equation}
Direct calculation gives for (\ref{G-a-cal})
\begin{equation}\label{G-a-cal-general}
\mathcal{G}_{mm'}\!\!=\!\!
\begin{pmatrix}
{\frac{1}{30}}(7 + \cos2\alpha) & -\frac{\sqrt{2}}{30}\sin2\alpha & -\frac{1}{15}\sin^2{\alpha} \\
-\frac{\sqrt{2}}{30}\sin2\alpha &  {\frac{1}{15}}(3 - \cos2\alpha) & 
-\frac{\sqrt{2}}{30}\sin2\alpha \\
-\frac{1}{15}\sin^2\alpha & -\frac{\sqrt{2}}{30}\sin2\alpha &  
{\frac{1}{30}}(7 + \cos2\alpha)
\end{pmatrix}\;.
\end{equation}

This expression simplifies for the magic angle 
$\alpha_{\mathrm{mag}}= \arccos{(1/\sqrt{3})}$:
\begin{eqnarray}\label{G-a-cal-magic}
\mathcal{G}_{mm'} = \frac{2}{9}
\left[{\widehat{\mathbb{I}}} - \frac{1}{5}
\left(
  \begin{array}{ccc}
    0 & 1 & 1 \\
    1 &  0 & 1 \\
    1 & 1 & 0 \\
  \end{array}
\right)
\right]
 \, ,
\end{eqnarray}
where ${\widehat{\mathbb{I}}}$ is the unit matrix.
Eqs.~(\ref{Gamma-a-expr}) and (\ref{G-a-cal-magic}) lead
to the above announced expressions (\ref{gamma-a}) and (\ref{R-j}).

\subsection*{2. Decay of a doubly excited state of the two-atom system}

Here we give more detail on the derivation of the results presented
in section \ref{fluo-2}. The decay of a doubly excited state, e.g.,
the state $|0, 0\rangle$ (i.e., $m_1=m_2=0$) of the two-atom system
is described by a quite complicated system of Schr\"{o}dinger
equations for the probability amplitudes entering Eq.~(\ref{Pci-2-t}).
However, the analysis of the system evolution can be simplified
due to the great difference $\varGamma_s \gg \varGamma_a$ between the
decay rates in the symmetric and antisymmetric channels, see
Fig.~\ref{Fig-decay}. This allows one to describe the time evolution
in the following succession of steps. At the first step, neglecting
the existence of the slow antisymmetric channel we find the time
evolution of the amplitude $A(t)$ in Eq.~(\ref{Pci-2-t}). At the
second step, using this fast decaying function $A(t)$ we calculate
a small probability of emission of a single antisymmetric photon.
Once occurred, this event determines further evolution of the
system via the antisymmetric channel.

{\em{Step 1. Decay rate via the symmetric channel.}} The time dependence
of $A(t)$ is mainly determined by the first transition of the
decay, see Fig.~\ref{Fig-decay}. Using the Hamiltonian
(\ref{H-int-2-b}), we obtain equations for the Laplace
transforms of the slow amplitudes
${\bar{A}}(t)=A(t)e^{2i\omega_{eg} t}$ and
$\bar{B}^s_{{\bf{k}} \lambda j}=B^s_{{\bf{k}} \lambda j} e^{2i\omega_{eg} t}$ as
\begin{multline}\label{Schr-Lapl-Psi-2}
 [z+ i(\omega_{{\bf{k}}}+E^{s}_j-2\omega_{eg})]\bar{B}^{s}_{{\bf{k}}\lambda j}[z]\\ =
 2i(g^{m=0,g}_{{\bf{k}}\lambda}C^{s}_{m=0,j})^* \bar{A}[z] + ... \;, \\
 z\bar{A}[z] = 1 + 2i\sideset{}{'}\sum_{{\bf{k}},\lambda, j}g^{m=0,g}_{{\bf{k}}\lambda}
 C^{s}_{m=0,j}\bar{B}^{s}_{{\bf{k}},\lambda,j}[z]
\; ,
\end{multline}
where we put $\cos({\bf{kr}}/2) \approx 1$. The omitted terms in the first
equation correspond to transitions from the singly excited symmetric states
(of energies $E^s_{j}$) to the ground state, they give only negligible corrections to
the studied decay rate of the doubly excited state. From the above system we express
$\bar{B}^s_{{\bf{k}} \lambda j }$ in terms of $\bar{A}[z]$, put it into the
last equation and perform a pole integration over $\omega_{{\bf{k}}}$.
In this way we arrive at the reduced equation for 
$\bar{A}[z]$:
\begin{equation}
[z+ \varGamma_{2s}/2]\tilde{A}[z]=1 \label{A(z)}\;,
\end{equation}
where
\begin{equation}
 \varGamma_{2s} = 4\omega^3_0 d^2_0
\sum_{j}C^{s}_{m=0,j}\!\left(C^{{s}}_{m'=0, j}\right)^*
\int'\frac{d\hat{o}}{4\pi} G_{0,0}(\theta,\varphi) \label{Gamma-2s-integral}
\; ,
\end{equation}
and the summation rule (\ref{sum-gg}) is used. The summation in $j$ gives
unity due to the unitarity of the matrix $\widehat{C}^s$.
Performing the angle integration over the upper hemisphere we find the expected relation
[see Eq.~(\ref{laplA0})]: $\varGamma_{2s}=\varGamma_{s}= 2\varGamma$.

{\em{Step 2. Leakage to the antisymmetric channel
[derivation of Eqs.~(\ref{Prob-definition})--(\ref{Prob})].}}
Now our aim is to find the probability amplitudes $B^a_{{\bf{k}} \lambda j}$
[see Eq.~(\ref{Pci-2-t})] of the antisymmetric states where the doubly
excited system may fall after the emission of an antisymmetric photon.
The short-time evolution of these amplitudes is mainly governed by
the antisymmetric part of the Hamiltonian (\ref{H-int-2-b}) that
connect these antisymmetric states with the doubly excited one
(see Fig.~\ref{Fig-decay}):
\begin{multline}\label{Schr-B-a}
\left[i{\frac{\partial}{\partial t}} - (E^a_j + \omega_{k})\right]B^{a}_{{\bf{k}}\lambda j}(t)\\ =
2i\left(g^{m=0,g}_{{\bf{k}}\lambda}C^{a}_{m=0,j}\right)^* \sin{\left(\frac{{\bf{kr}}}{2}\right)} A(t)
\; ,
\end{multline}
where the amplitude $A(t)$  is determined by the symmetric decay channel 
as in Eq.~(\ref{laplA0}).
The solution of  Eq.~(\ref{Schr-B-a}) 
with the initial condition $B^{a}_{{\bf{k}}\lambda j}(t=0)=0$ is given by
\begin{multline}\label{B-a-t}
B^{a}_{{\bf{k}}\lambda j}(t) =
 -{\frac{
 2i\left(g^{m=0,g}_{{\bf{k}}\lambda}C^{a}_{m=0,j}\right)^* 
 \sin\left({\frac{{\bf{kr}}}{2}}
 \right)}
{E^a_j + \omega_{{\bf{k}}} - 2\omega_{eg} +i\varGamma}}
 \\
 \times \left[e^{-(2i\omega_{eg} +\varGamma)t} - e^{-i(E^a_j + \omega_{{\bf{k}}})t}\right]
\; .
\end{multline}
When time $t$ is much greater than $1/\varGamma$ (but is still much shorter than the antisymmetric
decay time $~1/\varGamma^a$), this expression reduces to Eq.~(\ref{B-a-in}).

Now we derive the diagonal elements of the initial density matrix
(\ref{rho-in}) of the atomic subsystem (after the emission of a single
antisymmetric photon):
\begin{multline}\label{P-j-derivation}
P_j = \langle \psi^a_j| \hat{\rho}_{\textrm{in}}| \psi^a_j\rangle = \sideset{}{'}
\sum_{{\bf{k}}, \lambda} \left|B^{a}_{{\bf{k}}\lambda j}
\left(\frac{1}{\varGamma} \ll t \lesssim \frac{1}{\varGamma^a}\right)\right|^2  \\ 
=\sideset{}{'}
\sum_{{\bf{k}}, \lambda}\frac{4|g^{m=0,g}_{{\bf{k}}\lambda}C^{a}_{m=0,j}|^2}
{(\omega_{{\bf{k}}}+ E^a_j - 2\omega_{eg})^2 +\varGamma^2}\sin^2
{\left(\frac{\bf{kr}}{2}\right)}
\, .
\end{multline}
Performing here the pole integration 
in $\omega_k$ and using the summation rule
(\ref{sum-gg}), we obtain
\begin{eqnarray}\label{P-j-intermediate}
P_j = \frac{3}{8}k^2_0 r^2 |C^a_{m=0,j}|^2 \mathcal{G}_{0,0}(\alpha)
\; ,
\end{eqnarray}
where the approximation $\sin{x} \approx x$ is used, the matrix
$\mathcal{G}_{m,m'}(\alpha)$ is given by Eq.~(\ref{G-a-cal-general})
for an arbitrary angle $\alpha$ and by Eq.~(\ref{G-a-cal-magic}) for the
magic angle. Using Eq.~(\ref{G-a-cal-magic}), we arrive at the
result for $P_j$, Eq.~(\ref{Prob-j}). Finally, the total probability
for the doubly excited system to follow the antisymmetric
decay channel is given by the summation of (\ref{P-j-intermediate})
in $j$. Due to the unitarity of the matrix $\widehat{C}^a$ we arrive at
\begin{eqnarray}\label{P-intermediate}
P= \sum_j P_j = \frac{3}{8}k^2_0 r^2 \mathcal{G}_{0,0}(\alpha)
\, .
\end{eqnarray}
In general, this probability depends on the angle $\alpha$ 
as $P \propto (3 - \cos2\alpha)$.
For the magic angle $\alpha_{mag}$ we arrive at the expression (\ref{Prob}).

\section*{Appendix B: Derivation of Eq.~(\ref{PsPa}) for the spectrum of spontaneous 
emission in a strong magnetic field}
\renewcommand{\theequation}{B\arabic{equation}}
\setcounter{equation}{0}
For the spontaneous decay in a strong magnetic field of a doubly excited state 
$|m_1=0,m_2=0\rangle$, the wave function (\ref{Pci-2-t}) can be reduced to  
\begin{widetext}
\begin{multline}{\label{3Psit}}
\varPsi(t) = A(t)|0,0\rangle \otimes |{\textsl{vac}}\rangle+\int                              
{\widetilde{B}}_{{s}}(\omega_{1s},t)|s_0\rangle\otimes
|{\textsl{1}}_{\omega_{1s}}\rangle\,d\omega_{1s}+
\int{\widetilde{B}}_{{a}}(\omega_{1a},t)|a_0\rangle\otimes
|{\textsl{1}}_{\omega_{1a}}\rangle\,d\omega_{1a}\\
+\int\!\!\!\int_{\omega_{1s}\leqslant \omega_{2s}}
{\widetilde{C}}_{s}(\omega_{1s},\omega_{2s},t)|g,g\rangle\otimes
|{\textsl{1}}_{\omega_{1s}}{\textsl{1}}_{\omega_{2s)}}\rangle\;d\omega_{1s}
d\omega_{2s}+\int\!\!\!\int_{\omega_{1a}\leqslant \omega_{2a}}
{\widetilde{C}}_a(\omega_{1a},\omega_{2a},t)|g,g\rangle\otimes
|{\textsl{1}}_{\omega_{1a}}{\textsl{1}}_{\omega_{2a}}\rangle\;d\omega_{1a}
d\omega_{2a}\;.
\end{multline}
\end{widetext}
Here the following designations are used: (i) the one-photon wave packets 
$|{\textsl{1}}_{\omega s}\rangle$ and $|{\textsl{1}}_{\omega a}\rangle$ 
are introduced being the unique superpositions of the corresponding symmetric and 
antisymmetric one-photon states $b^\dag({\bf{k}})|{\textsl{vac}}\rangle$ 
with different ${\bf{k}}$ (and the same $|{\bf{k}}|$), that alone interact with the state 
$m=0$ \cite{FANO_MAK}; 
(ii) $|{\textsl{1}}_{\omega_{1s}}{\textsl{1}}_{\omega_{2s}}\rangle$ and
$|{\textsl{1}}_{\omega_{1a}}{\textsl{1}}_{\omega_{2a}}\rangle$ are two-photon states 
where $\omega_{1\nu}\leqslant\omega_{2\nu}$ is taken for definiteness; 
(iii) and indexes are simplified for the probability amplitudes ${\widetilde{B}}$ and 
${\widetilde{C}}$ that replace $B$ and $C$; (iv) a shorter notation is used for the singly 
excited states, $|s_0=\left|\psi_{m=0}^{(\nu=s)}\right\rangle$ and 
$|a_0\rangle=\left|\psi_{m=0}^{(\nu=a)}\right\rangle$  

The Shr\"odinger equation for the wave function (\ref{3Psit}) is given by the following set 
of equations for the probability amplitudes (where $\varDelta$ is the splitting between 
the states $|s_0\rangle$ and $|a_0\rangle$):
\begin{widetext}
\begin{equation}{\label{dABCdt}}
\begin{split}
&{\frac{dA}{dt}}=-2i\omega_{eg}A
-i\sqrt{{\frac{\varGamma_{{s}}}{2\pi}}}
\int\limits_{-\infty}^{\infty}{\widetilde{B}}_{s}(\xi)\,d\xi-i\sqrt{{\frac{\varGamma_{{a}}}{2\pi}}}
\int\limits_{-\infty}^{\infty}{\widetilde{B}}_{a}(\xi)\,d\xi\;,\\
&{\frac{d{\widetilde{B}}_{s}(\omega_{1s})}{dt}}=
-i\left(\omega_{eg}+{\frac{1}{2}}\varDelta+\omega_{1s}\right){\widetilde{B}}_s
-i\sqrt{{\frac{\varGamma_{{s}}}{2\pi}}}A-
i\sqrt{{\frac{\varGamma_{{s}}}{2\pi}}}
\left(\int\limits_{-\infty}^{\omega_{1s}}{\widetilde{C}}_s(\xi,\omega_{1s})\,d\xi+
\int\limits_{\omega_{1s}}^{\infty}{\widetilde{C}}_s(\omega_{1s},\xi)\;d\xi\right)\;,\\
&{\frac{d{\widetilde{B}}_{a}(\omega_{1a})}{dt}}=-i\left(\omega_{eg}-{\frac{1}{2}}\varDelta+\omega_{1a}\right){\widetilde{B}}_{{a}}
-i\sqrt{{\frac{\varGamma_{{a}}}{2\pi}}}A-
i\sqrt{{\frac{\varGamma_{{a}}}{2\pi}}}
\left(\int\limits_{-\infty}^{\omega_{1a}}{\widetilde{C}}_a(\xi,\omega_{1a})\,d\xi+
\int\limits_{\omega_{1a}}^{\infty}{\widetilde{C}}_a(\omega_{1a},\xi)\;d\xi\right)\;,\\
&{\frac{d{\widetilde{C}}_{s}(\omega_{1s},\omega_{{2s}})}{dt}}=-i(\omega_{1s}+\omega_{2s}){\widetilde{C}}_{s}
-
-i\sqrt{{\frac{\varGamma_s}{2\pi}}}
\left[{\widetilde{B}}_{s}(\omega_{1s})+{\widetilde{B}}_{s}(\omega_{2s})\right]]\;,\\
&{\frac{d{\widetilde{C}}_{a}(\omega_{1a},\omega_{{2a}})}{dt}}=-i(\omega_{1a}+\omega_{2a}){\widetilde{C}}_{a}
-
-i\sqrt{{\frac{\varGamma_a}{2\pi}}}
\left[{\widetilde{B}}_{a}(\omega_{1a})+{\widetilde{B}}_{a}(\omega_{2a})\right]\;. 
\end{split}
\end{equation}
\end{widetext}
These equations incorporate the idealizations introduced by the Weisskopf--Wigner approximation \cite{WW}: integrations in photon frequencies are performed from $-\infty$ to $+\infty$; non-diagonal matrix elements 
of the atom--field interaction do not depend on $\omega$, so, being expressed in terms of  
the spontaneous decay rates $\varGamma_s$ and $\varGamma_a$ 
[see Eqs. (\ref{gamma-s}) and (\ref{gamma-a-H})], they are pulled out the corresponding 
integrals.

The initial condition of interest is
\begin{equation}{\label{ABCt0}}
\begin{split}
A(t=0)=1\;,\;\;
{\widetilde{B}}_{s}(t=0)={\widetilde{B}}_{a}(t=0)&=0\;,\\
{\widetilde{C}}_{s}(t=0)={\widetilde{C}}_{a}(t=0)&=0\;.
\end{split}
\end{equation}
The solution of Eq.~(\ref{dABCdt}) with the initial condition (\ref{ABCt0}) can be obtained by 
the method described in Ref.~\cite{MAK_YUD}. It is convenient to represent the solution in the following integral form that is easy to check:
\begin{widetext}
\begin{equation}{\label{ABCt}}
\begin{split}
A=&
{\frac{1}{2\pi i}}\int\limits_{-\infty}^{\infty}{\frac{e^{izt}\,dz}{
z+2\omega_{eg}-{\frac{i}{2}}(\varGamma_s+\varGamma_a)}}\;,\\
{\widetilde{B}}_{s}=&-{\frac{1}{2\pi i}}{\sqrt{{\frac{\varGamma_s}{2\pi}}}}
\int\limits_{-\infty}^{\infty}
{\frac{e^{izt}\,dz}{\left[z+2\omega_{eg}-{\frac{i}{2}}(\varGamma_s+\varGamma_a)\right]\left(
z+\omega_{eg}+{\frac{1}{2}}\varDelta-{\frac{i}{2}}\varGamma_s+\omega_{1s}\right)}}\;,\\
{\widetilde{B}}_{a}=&-{\frac{1}{2\pi i}}{\sqrt{{\frac{\varGamma_a}{2\pi}}}}
\int\limits_{-\infty}^{\infty}
{\frac{e^{izt}\,dz}{\left[z+2\omega_{eg}-{\frac{i}{2}}(\varGamma_s+\varGamma_a)\right]\left(
z+2\omega_{eg}-{\frac{1}{2}}\varDelta-{\frac{i}{2}}\varGamma_a+\omega_{1a}\right)}}\;,\\
{\widetilde{C}}_{s}=&{\frac{\varGamma_s}{4\pi^2i}}\left\{\int\limits_{-\infty}^{\infty}
{\frac{e^{izt}\,dz}{\left[z+2\omega_{eg}-{\frac{i}{2}}(\varGamma_s+\varGamma_a)\right]
\left(z+\omega_{eg}+{\frac{1}{2}}\varDelta-{\frac{i}{2}}\varGamma_s+\omega_{1s}\right)
(z+\omega_{1s}+\omega_{2s})}}\right.\;,\\
&+\left.\int\limits_{-\infty}^{\infty}{\frac{e^{izt}\,
dz}{\left[z+2\omega_{eg}-{\frac{i}{2}}(\varGamma_s+\varGamma_a)\right]\left(
z+\omega_{eg}+{\frac{1}{2}}\varDelta-{\frac{i}{2}}\varGamma_s+\omega_{2s}\right)
(z+\omega_{1s}+\omega_{2s})}}\right\}\\
{\widetilde{C}}_{a}=&{\frac{\varGamma_a}{4\pi^2i}}\left\{\int\limits_{-\infty}^{\infty}{\frac{e^{izt}\,dz}{\left[z+2\omega_{eg}-{\frac{i}{2}}(\varGamma_s+\varGamma_a)\right]\left(
z+\omega_{eg}-{\frac{1}{2}}\varDelta-{\frac{i}{2}}\varGamma_a+\omega_{1a}\right)
(z+\omega_{1a}+\omega_{2a})}}\right.\\
&+\left.\int\limits_{-\infty}^{\infty}{\frac{e^{izt}\,
dz}{\left[z+2\omega_{eg}-{\frac{i}{2}}(\varGamma_s+\varGamma_a)\right]\left(
z+\omega_{eg}-{\frac{1}{2}}\varDelta-{\frac{i}{2}}\varGamma_a+\omega_{2a}\right)
(z+\omega_{1a}+\omega_{2a})}}\right\}\;.
\end{split}
\end{equation}
The emission spectrum is defined as 
\begin{equation}{\label{spectrum-definition}}
{\mathcal{S}}(\omega)=\int\limits_{-\infty}^{\infty}\left|{\widetilde{C}}_{s}(\omega,\xi;t
\rightarrow\infty)\right|^2\,d\xi+
\int\limits_{-\infty}^{\infty}\left|{\widetilde{C}}_{a}(\omega,\xi;t
\rightarrow\infty)\right|^2\,d\xi\;.
\end{equation}
An intermediate calculation gives 
\begin{equation}{\label{Ctinfty}}
\begin{split}
&{\widetilde{C}}_{s}(\omega_{1s},\omega_{2s};t\rightarrow\infty)={\frac{\varGamma_s}{2\pi}}\cdot
{\frac{\exp[-i(\omega_{1s}+\omega_{2s})t]}
{\omega_{1s}+\omega_{2s}-2\omega_{eg}+{\frac{i}{2}}(\varGamma_s+\varGamma_a)}}\left(
{\frac{1}{\omega_{1s}-\omega_{eg}-{\frac{1}{2}}\varDelta+{\frac{i}{2}}\varGamma_s}}+{\frac{1}
{\omega_{2s}-\omega_{eg}-{\frac{1}{2}}\varDelta+{\frac{i}{2}}\varGamma_s}}\right)\;,\\
&{\widetilde{C}}_{a}(\omega_{1a},\omega_{2a};t\rightarrow\infty)={\frac{\varGamma_a}{2\pi}}\cdot
{\frac{\exp[-i(\omega_{1a}+\omega_{2a})t]}
{\omega_{1a}+\omega_{2a}-2\omega_{eg}+{\frac{i}{2}}(\varGamma_s+\varGamma_a)}}\left(
{\frac{1}{\omega_{1a}-\omega_{eg}+{\frac{1}{2}}\varDelta+{\frac{i}{2}}\varGamma_a}}+{\frac{1}
{\omega_{2a}-\omega_{eg}+{\frac{1}{2}}\varDelta+{\frac{i}{2}}\varGamma_a}}\right)\;.
\end{split}
\end{equation}
Hence, one arrives at
\begin{multline}
{\mathcal{S}}(\omega)={\frac{\varGamma_s^2}{4\pi^2}}{\frac{1}{(\omega-\omega_{eg}-
{\frac{1}{2}}\varDelta)^2+{\frac{1}{4}}\varGamma_s^2}}
\int\limits_{-\infty}^{\infty}
{\frac{(\xi+\omega-2\omega_{eg}-\varDelta)^2+\varGamma_s^2}{
\left[(\xi-\omega_{eg}-{\frac{1}{2}}\varDelta)^2+{\frac{1}{4}}\varGamma_s^2\right]
\left[(\xi+\omega-2\omega_{eg})^2+{\frac{1}{4}}(\varGamma_s+\varGamma_a)^2\right]}}\,d\xi\\
+{\frac{\varGamma_a^2}{4\pi^2}}{\frac{1}{(\omega-\omega_{eg}+
{\frac{1}{2}}\varDelta)^2+{\frac{1}{4}}\varGamma_a^2}}
\int\limits_{-\infty}^{\infty}
{\frac{(\xi+\omega-2\omega_{eg}+\varDelta)^2+\varGamma_s^2}{
\left[(\xi-\omega_{eg}+{\frac{1}{2}}\varDelta)^2+{\frac{1}{4}}\varGamma_a^2\right]
\left[(\xi+\omega-2\omega_{eg})^2+{\frac{1}{4}}(\varGamma_s+\varGamma_a)^2\right]}}\,d\xi\;.
\end{multline}
Finally, performing the integration, one gets Eq.~(\ref{PsPa}) that have been used for 
calculation of the spectra shown in Fig.~\ref{spectrum-strong-H}.
\end{widetext}  
\section*{Appendix C: Derivation of Eq. (\ref{l-direction})} 
\renewcommand{\theequation}{C\arabic{equation}}
\setcounter{equation}{0}
As described in Sec.~\ref{angular-resolution}, to obtain the unit vector 
$\widehat{\mathbf{r}}$
in the direction of the vector ${\bf{r}}$ connecting the two 
atoms, one should
find any two nonparallel directions $\widehat{\mathbf{h}}_1$ and $\widehat{\mathbf{h}}_2$ of the magnetic
field such that the spectral detuning between the symmetric and antisymmetric emission
peaks is minimal. This condition means that the scalar products
$\widehat{\mathbf{r}}\widehat{\mathbf{h}}_1=\widehat{\mathbf{r}}\widehat{\mathbf{h}}_2 =
\cos{(\alpha_{\textrm{mag}})}$. The unit vector
$\hat{\mathbf{r}}$ can be expanded over the (non-orthogonal) basis of three vectors
\begin{eqnarray}\label{l-expantion}
\widehat{\mathbf{r}} = a \widehat{\mathbf{h}}_1 + b \widehat{\mathbf{h}}_2 +
c [\widehat{\mathbf{h}}_1 \times \widehat{\mathbf{h}}_2]
 \; ,
\end{eqnarray}
where $[\hat{\mathbf{h}}_1 \times \hat{\mathbf{h}}_2]$ means the vector product.
The coefficients $a$ and $b$ can be obtained from two relations:
\begin{equation}{\label{cos}}
\begin{split}
&\cos{(\alpha_{\textrm{mag}})}
= \widehat{\mathbf{r}}\widehat{\mathbf{h}}_1 = a +
b \widehat{\mathbf{h}}_1 \widehat{\mathbf{h}}_2\;,\\
&\cos{(\alpha_{\textrm{mag}})} = \widehat{\mathbf{r}}\widehat{\mathbf{h}}_2 =
a\widehat{\mathbf{h}}_1\widehat{\mathbf{h}}_2 + b\;.
\end{split}
\end{equation}
Hence,
\begin{equation}
a=b = \frac{\cos{(\alpha_{\textrm{mag}})}}{1 + \cos{(\vartheta_{12})}}
\end{equation}
where $\vartheta_{12}$ is the angle between $\widehat{\mathbf{h}}_1$ and
$\widehat{\mathbf{h}}_2$.
The coefficient $c$ in the expansion (\ref{l-expantion}) is obtained
(up to the sign) from the relation $1 = \widehat{\mathbf{r}}^2 =
a^2(\widehat{\mathbf{h}}_1 + \widehat{\mathbf{h}}_2)^2 +
c^2 [\widehat{\mathbf{h}}_1 \times \widehat{\mathbf{h}}_2]$. Hence,
\begin{eqnarray}\label{c}
c = \pm \frac{1}{\sin{\vartheta_{12}}}\sqrt{1 - \frac{2\cos^2{(\alpha_{\textrm{mag}})}}{1+\cos{\vartheta_{12}}}}
[\widehat{\mathbf{h}}_1 \times \widehat{\mathbf{h}}_2 ]
 \, .
\end{eqnarray}
Collecting Eqs.~(\ref{l-expantion})--(\ref{c}) we arrive at
Eq.~(\ref{l-direction}) presented in the section \ref{angular-resolution}.

\newpage\clearpage

\begin{thebibliography}{99}
\bibitem{CONTROL} The same configuration of two atoms in the magnetic field 
was considered in Ref.~\cite{JETPLETT} where we suggested a method to create
subradiant states with about 95\% fidelity. This method is based 
on a proper manipulation of the magnetic field. 
However, spectroscopic aspects of the system were not considered in Ref.~\cite{JETPLETT}.
\end{thebibliography}
\end{document}